%% file: id-bim-dust.tex
\documentclass[11pt,a4paper,english]{article}
\pdfoutput=1

\usepackage{lmodern}
\renewcommand{\sfdefault}{lmss}

\usepackage{color}
\usepackage{float}
\usepackage{mathtools}
\usepackage{enumitem}
\usepackage{amsmath}
\usepackage{amsthm}
\usepackage{amssymb}
\usepackage{graphicx}

\makeatletter

\input{preamble}

\makeatother

\usepackage{babel}
\begin{document}

\input{macros.tex}

\newcommand{\panL}[1]{\raisebox{3ex}{\footnotesize\bf(#1)}\hspace{-5mm}}


\newcommand{\myReleaseInfo}{~}

\subheader{spherical dust collapse in bimetric relativity}

\newcommand{\myTitle}{Bimetric polytropes}

\newcommand{\myAbstract}{We present a method for solving the constraint
equations in the Hassan\textendash Rosen bimetric theory to determine
the initial data for the gravitational collapse of spherically symmetric
dust. The setup leads to equations similar to those for a polytropic
fluid in general relativity, here called a generalized Lane\textendash Emden
equation. Using a numerical code which solves the evolution equations
in the standard 3+1 form, we also obtain a short term development
of the initial data for these bimetric polytropes. The evolution highlights
some important features of the bimetric theory such as the interwoven
and oscillating null cones representing the essential nonbidiagonality
in the dynamics of the two metrics. The simulations are in the strong-field
regime and show that, at least at an early stage, the collapse of
a dust cloud is similar to that in general relativity, and with no
instabilities, albeit with small oscillations in the metric fields.}

\newcommand{\myKeywords}{\bgroup\small Modified gravity, Ghost-free
bimetric theory, Numerical bimetric relativity\egroup}

\title{\myTitle}

\author{Mikica Kocic,}

\author{Francesco Torsello,}

\author{Marcus H\"{o}g\r{a}s,}

\author{and Edvard M\"{o}rtsell}

\affiliation{%
  Department of Physics \& The Oskar Klein Centre,\\
  Stockholm University, AlbaNova University Centre,
  SE-106 91 Stockholm
}

\email{mikica.kocic@fysik.su.se}

\email{francesco.torsello@fysik.su.se}

\email{marcus.hogas@fysik.su.se}

\email{edvard@fysik.su.se}

\hypersetup{
  pdftitle=\myTitle,
  pdfauthor=Mikica Kocic et al.,
  pdfsubject=Hassan-Rosen ghost-free bimetric theory,
  pdfkeywords={Modified gravity, Ghost-free bimetric theory,
    Numerical bimetric relativity}
}

\renewcommand\afterTocSpace{\vspace{3ex}} 
\addtocontents{toc}{\vspace{-3ex}}
\emitFrontMatter


\input{sec-10.tex}\vfill

\input{sec-20.tex}

\input{sec-22.tex}

\input{sec-23.tex}

\input{sec-30.tex}

\input{sec-33.tex}

\input{sec-60.tex}


\emitAppendix

\input{sec-90.tex}


\ifprstyle

\bibliographystyle{apsrev4-1}
\bibliography{id-bim-dust}

\else

\bibliographystyle{JHEP}
\bibliography{id-bim-dust}

\fi
\end{document}

%% file: preamble.tex
\newif\ifprstyle

\ifx\affiliation\undefined
  \prstylefalse
\else
  \prstyletrue
  \newif\ifnotoc
  \notoctrue
\fi

\ifprstyle
  \usepackage[
    colorlinks=true, pdfa=true,
    urlcolor=blue,   anchorcolor=blue,
    citecolor=blue,  filecolor=blue,
    linkcolor=blue,  menucolor=blue,
    pagecolor=blue,  linktocpage=true
  ]{hyperref}
\else
  \usepackage{jheppub}
  \newcommand{\email}[1]{\emailAdd{#1}}
\fi

\ifx\ifCrop\undefined\else

  \ifx\ifPresentation\undefined
     \usepackage[paperwidth=171.2mm,paperheight=261.2mm]{geometry}
  \else
     \usepackage[paperwidth=330mm,paperheight=233mm]{geometry}
  \fi

  \divide\@tempdima\baselineskip
  \@tempcnta=\@tempdima
  \ifx\ifPresentation\undefined
    \hypersetup{pdfstartview={Fit},pdfpagelayout={TwoPageRight}}
  \else
    \hypersetup{pdfstartview={Fit},pdfpagelayout={SinglePage}}

    \usepackage{everypage}
    \AddEverypageHook{\tikz[remember picture,overlay]{%
      \draw [gray!10] ($(current page.east)+(-130mm,+180mm)$)
         -- ($(current page.east)+(-130mm,-100mm)$);}}
  \fi
\fi

\hypersetup{
    urlcolor=black!40!blue,   anchorcolor=black!40!blue,
    citecolor=black!40!blue,  filecolor=black!40!blue,
    linkcolor=black!40!blue,  menucolor=black!40!blue,
    pagecolor=black!40!blue
}

\usepackage{bookmark}

\usepackage{amsmath,amsfonts,amssymb,bm,cancel}

\usepackage{color}
\usepackage{colortbl}

\ifprstyle
  
\else
  
\fi

\newcommand{\bSe}{\begin{subequations}}
\newcommand{\eSe}{\end{subequations}}

\newcommand{\bWe}{\begin{widetext}}
\newcommand{\eWe}{\end{widetext}}

\allowdisplaybreaks

\usepackage{tikz}
\usetikzlibrary{shapes,arrows}
\usetikzlibrary{calc}
\usetikzlibrary{positioning}
\usetikzlibrary{decorations.pathreplacing}
\usetikzlibrary{shapes.misc}

\usepackage[scr,scaled=1.1]{rsfso}
\usepackage{eucal}

\DeclareMathAlphabet{\mathsfit}{\encodingdefault}{\sfdefault}{m}{sl}
\SetMathAlphabet{\mathsfit}{bold}{\encodingdefault}{\sfdefault}{bx}{sl}

\renewcommand\floatc@plain[2]{\setbox\@tempboxa\hbox{{\@fs@cfont #1.} #2}%
\ifdim\wd\@tempboxa>\hsize {\@fs@cfont #1.} #2\par
\else\hbox to\hsize{\hfil\box\@tempboxa\hfil}\fi}

\ifprstyle\else
\fi

\newcommand{\emitFrontMatter}{
  \ifprstyle
    \begin{abstract}\myAbstract\end{abstract}
  \else
    \ifnotoc
      \abstract{\\[-1.64ex]\hphantom{\hspace{0.28cm}}%
        \parbox{1\columnwidth-0.28cm}%
        {\renewcommand\baselinestretch{1.05}\small\hspace{18mm}\myAbstract}}
    \else
      \abstract{\myAbstract}
    \fi
  \fi

  \keywords{\myKeywords}

  \ifprstyle\else
    \makeatletter
      \def\@fpheader{\myReleaseInfo}
    \makeatother
    \ifnotoc
      \compress
      \renewcommand\afterLogoSpace{}
      \renewcommand\afterSubheaderSpace{}
      \renewcommand\afterProceedingsSpace{}
      \makeatletter
      \renewcommand\ps@titlepage{}
      \makeatother
      \toccontinuoustrue
    \else
      \addtocontents{toc}{\protect\setcounter{tocdepth}{2}}
    \fi
  \fi

  \maketitle

  \ifprstyle\else
    \ifnotoc
      \hrule\bigskip\bigskip
    \else
      \flushbottom
    \fi
  \fi
}

\usepackage{titlesec}

\newcommand{\emitAppendix}{
  \appendix
}



%% file: macros.tex
\newif\ifColors

\newif\ifShowSigns 

\ifx \ii \undefined

\definecolor{red}{rgb}{1,0,0.1}
\definecolor{green}{rgb}{0.0,0.6,0}
\definecolor{blue}{rgb}{0.1,0.1,1}
\definecolor{brown}{rgb}{0.6,0.3,0}
\definecolor{orange}{rgb}{0.8,0.3,0}
\definecolor{magenta}{rgb}{0.9,0.1,1}

\newcommand\nPlusOne{$N$+1}

\renewcommand\tilde[1]{\mkern1mu\widetilde{\mkern-1mu#1}}

\global\long\def\ii{\mathrm{i}}
\global\long\def\ee{\mathrm{e}}
\global\long\def\dd{\mathrm{d}}
\global\long\def\ppi{\mathrm{\pi}}
\global\long\def\tr{\mathsf{{\scriptscriptstyle T}}}
\global\long\def\Tr{\operatorname{Tr}}
\global\long\def\op#1{\operatorname{#1}}
\global\long\def\dim{\operatorname{dim}}
\global\long\def\diag{\operatorname{diag}}
\global\long\def\Lie{\mathrm{\mathscr{L}}}

\global\long\def\mfrac#1#2{\frac{\raisebox{-0.45ex}{\scalebox{0.9}{#1}}}{\raisebox{0.4ex}{\scalebox{0.9}{#2}}}}
\global\long\def\mbinom#1#2{\Big(\begin{array}{c}
 #1\\[-0.75ex]
 #2 
\end{array}\Big)}

\global\long\def\tud#1#2#3{#1{}^{#2}{}_{#3}}
\global\long\def\tdu#1#2#3{#1{}_{#2}{}^{#3}}

\global\long\def\qvf{\xi}
\global\long\def\ixA{a}
\global\long\def\ixB{b}
\global\long\def\ccVar{\mathcal{C}}

\global\long\def\lidx#1{\ ^{(#1)}\!}

\global\long\def\gSector#1{{\color{black}#1}}
\global\long\def\fSector#1{{\color{black}#1}}
\global\long\def\hSector#1{{\color{black}#1}}
\global\long\def\sSector#1{{\color{black}#1}}
\global\long\def\lSector#1{{\color{black}#1}}
\global\long\def\mSector#1{{\color{black}#1}}
\global\long\def\hrColor#1{{\color{black}#1}}
\global\long\def\VColor#1{{\color{black}#1}}
\global\long\def\KColor#1{{\color{black}#1}}
\global\long\def\KVColor#1{{\color{black}#1}}
\global\long\def\pfSector#1{{\color{black}#1}}

\global\long\def\gMet{\gSector g}
\global\long\def\gSp{\gSector{\gamma}}
\global\long\def\gK{\gSector K}
\global\long\def\gE{\gSector e}
\global\long\def\gD{\gSector D}
\global\long\def\gR{\gSector R}
\global\long\def\gCS{\gSector{\Gamma}}
\global\long\def\gVse{\gSector{V_{g}}}
\global\long\def\gTse{\gSector{T_{g}}}
\global\long\def\gEinst{\gSector{G_{g}}}
\global\long\def\gRicci{\gSector{R_{g}}}
\global\long\def\gCC{\gSector{\mathcal{C}}}
\global\long\def\gCE{\gSector{\mathcal{E}}}
\global\long\def\gCD{\gSector{\nabla}}
\global\long\def\gPi{\gSector{\pi}}

\global\long\def\fMet{\fSector f}
\global\long\def\fSp{\fSector{\varphi}}
\global\long\def\fK{\fSector{\tilde{K}}}
\global\long\def\fE{\fSector m}
\global\long\def\fD{\fSector{\tilde{D}}}
\global\long\def\fR{\fSector{\tilde{R}}}
\global\long\def\fCS{\fSector{\tilde{\Gamma}}}
\global\long\def\fVse{\fSector{V_{f}}}
\global\long\def\fTse{\fSector{T_{f}}}
\global\long\def\fEinst{\fSector{G_{f}}}
\global\long\def\fRicci{\fSector{R_{f}}}
\global\long\def\fCC{\fSector{\widetilde{\mathcal{C}}}}
\global\long\def\fCE{\fSector{\widetilde{\mathcal{E}}}}
\global\long\def\fCD{\fSector{\widetilde{\nabla}}}
\global\long\def\fPi{\fSector p}
\global\long\def\fPi{\fSector{\tilde{\pi}}}

\global\long\def\gLapse{\gSector{\alpha}}
\global\long\def\gShift{\gSector{\beta}}
\global\long\def\gShiftVec{\gSector{\beta}}

\global\long\def\fLapse{\fSector{\tilde{\alpha}}}
\global\long\def\fShift{\fSector{\tilde{\beta}}}
\global\long\def\fShiftVec{\fSector{\tilde{\beta}}}

\global\long\def\gKappa{\gSector{\kappa_{g}}}
\global\long\def\gKappainv{\gSector{\kappa_{g}^{-1}}}
\global\long\def\Mg{\gSector{M_{g}^{d-2}}}

\global\long\def\fKappa{\fSector{\kappa_{f}}}
\global\long\def\fKappainv{\fSector{\kappa_{f}^{-1}}}
\global\long\def\Mf{\fSector{M_{f}^{d-2}}}

\global\long\def\grho{\gSector{\rho}}
\global\long\def\gjota{\gSector j}
\global\long\def\gJota{\gSector J}

\global\long\def\frho{\fSector{\tilde{\rho}}}
\global\long\def\fjota{\fSector{\tilde{j}}}
\global\long\def\fJota{\fSector{\tilde{J}}}

\global\long\def\grhom{\gSector{\rho^{\mathrm{m}}}}
\global\long\def\gjotam{\gSector{j^{\mathrm{m}}}}
\global\long\def\gJotam#1{\gSector{J_{#1}^{\mathrm{m}}}}

\global\long\def\frhom{\fSector{\tilde{\rho}^{\mathrm{m}}}}
\global\long\def\fjotam{\fSector{\tilde{j}^{\mathrm{m}}}}
\global\long\def\fJotam#1{\fSector{\tilde{J}_{#1}^{\mathrm{m}}}}

\global\long\def\grhob{\gSector{\rho^{\mathrm{b}}}}
 \global\long\def\gjotab{\gSector{j^{\mathrm{b}}}}
 \global\long\def\gJotab{\gSector{J^{\mathrm{b}}}}

\global\long\def\frhob{\fSector{\tilde{\rho}^{\mathrm{b}}}}
 \global\long\def\fjotab{\fSector{\tilde{j}^{\mathrm{b}}}}
 \global\long\def\fJotab{\fSector{\tilde{J}^{\mathrm{b}}}}

\global\long\def\grhoeff{\gSector{\rho_{\mathrm{eff}}}}
 \global\long\def\gjotaeff{\gSector{j_{\mathrm{eff}}}}
 \global\long\def\gJotaeff{\gSector{J_{\mathrm{eff}}}}

\global\long\def\frhoeff{\fSector{\tilde{\rho}_{\mathrm{eff}}}}
 \global\long\def\fjotaeff{\fSector{\tilde{j}_{\mathrm{eff}}}}
 \global\long\def\fJotaeff{\fSector{\tilde{J}_{\mathrm{eff}}}}

\global\long\def\gAlpha{\gSector{\alpha}}
\global\long\def\gBeta{\gSector{\beta}}
\global\long\def\gEA{\gSector A}
\global\long\def\gEB{\gSector B}
\global\long\def\fAlpha{\fSector{\tilde{\alpha}}}
\global\long\def\fBeta{\fSector{\tilde{\beta}}}
\global\long\def\fEA{\fSector{\tilde{A}}}
\global\long\def\fEB{\fSector{\tilde{B}}}

\global\long\def\sEtau{\mSector{\tau}}
\global\long\def\sESigma{\mSector{\Sigma}}
\global\long\def\sER{\mSector R}

\global\long\def\cphi{\gSector{\psi}}
\global\long\def\gcphi{\gSector{\psi}_{\gMet}}
\global\long\def\fcphi{\fSector{\psi}_{\fMet}}

\global\long\def\gblab{\gSector{\mathrm{b}}}
\global\long\def\gmlab{\gSector{\mathrm{m}}}
\global\long\def\fblab{\fSector{\mathrm{b}}}
\global\long\def\fmlab{\fSector{\mathrm{m}}}

\global\long\def\pfrho{\pfSector{\rho}_{\pfSector 0}}
\global\long\def\pfD{\pfSector{\hat{D}}}
\global\long\def\pfS{\pfSector{\hat{S}}}
\global\long\def\pftau{\pfSector{\hat{\tau}}}
\global\long\def\pfu{\pfSector u}
\global\long\def\pfv{\pfSector{\hat{v}}}
\global\long\def\pfW{\pfSector w}
\global\long\def\pfh{\pfSector h}
\global\long\def\pfeps{\pfSector{\epsilon}}
\global\long\def\pfP{\pfSector P}

\global\long\def\Proj{\operatorname{\perp}}
\global\long\def\gProj{\gSector{\operatorname{\perp}_{g}}}
\global\long\def\fProj{\fSector{\operatorname{\perp}_{f}}}
\global\long\def\hProj{\hSector{\operatorname{\perp}}}
\global\long\def\prho{\boldsymbol{\rho}}
\global\long\def\pjota{\boldsymbol{j}}
\global\long\def\pJota{\boldsymbol{J}}

\global\long\def\sgn{\gSector{\mathsfit{n}{\mkern1mu}}}
\global\long\def\sgD{\gSector{\mathcal{D}}}
\global\long\def\sgQ{\gSector{\mathcal{Q}}}
\global\long\def\sgV{\gSector{\mathcal{V}}}
\global\long\def\sgU{\gSector{\mathcal{U}}}
\global\long\def\sgB{\gSector{\mathcal{B}}}

\global\long\def\sfn{\fSector{\tilde{\mathsfit{n}}{\mkern1mu}}}
\global\long\def\sfD{\fSector{\widetilde{\mathcal{D}}}}
\global\long\def\sfQ{\fSector{\widetilde{\mathcal{Q}}}}
\global\long\def\sfV{\fSector{\widetilde{\mathcal{V}}}}
\global\long\def\sfU{\fSector{\widetilde{\mathcal{U}}}}
\global\long\def\sfB{\fSector{\widetilde{\mathcal{B}}}}

\global\long\def\sgW{\gSector{\mathcal{W}}}
\global\long\def\sgQU{\gSector{(\mathcal{Q\fSector{{\scriptstyle \widetilde{U}}}})}}

\global\long\def\sfW{\fSector{\tilde{\mathcal{W}}}}
\global\long\def\sfQU{\fSector{(\mathcal{\widetilde{Q}\gSector{{\scriptstyle U}}})}}

\global\long\def\hMet{\hSector h}
\global\long\def\hSp{\hSector{\chi}}
\global\long\def\hLapse{\hSector H}
\global\long\def\hShift{\hSector q}
\global\long\def\hShiftVec{\hSector q}
\global\long\def\hCC{\hSector{\bar{\mathcal{C}}}}

\global\long\def\sLs{\lSector{\hat{\Lambda}}}
\global\long\def\sLt{\lSector{\lambda}}
\global\long\def\sLtinv{\lSector{\lambda^{-1}}}
\global\long\def\sLv{\lSector v}
\global\long\def\sLp{\lSector p}
\global\long\def\sRs{\lSector{\hat{R}}}
\global\long\def\sRbar{\lSector{\bar{R}}}

\global\long\def\sI{\lSector{\hat{I}}}
\global\long\def\sEta{\lSector{\hat{\delta}}}

\global\long\def\betap#1{\beta_{{\scriptscriptstyle (#1)}}}
\global\long\def\betaScale{\ell^{-2}}
\global\long\def\betaSum{\betaScale{\textstyle \sum_{n}}\beta_{(n)}}
\global\long\def\betaSumL{\betaScale{\displaystyle \sum_{n=0}^{4}}\beta_{(n)}}

\global\long\def\signV{\,+\,}
\global\long\def\isignV{\,-\,}
\global\long\def\usignV{}
\global\long\def\uisignV{-\,}

\global\long\def\isignV{\,+\,}
\global\long\def\signV{\,-\,}
\global\long\def\uisignV{}
\global\long\def\usignV{-\,}

\global\long\def\signK{\,+\,}
\global\long\def\isignK{\,-\,}
\global\long\def\usignK{}
\global\long\def\uisignK{-\,}

\global\long\def\isignK{\,+\,}
\global\long\def\signK{\,-\,}
\global\long\def\uisignK{}
\global\long\def\usignK{-\,}

\global\long\def\signKV{\,+\,}
\global\long\def\isignKV{\,-\,}
\global\long\def\usignKV{}
\global\long\def\uisignKV{-\,}

\global\long\def\isignKV{\,+\,}
\global\long\def\signKV{\,-\,}
\global\long\def\uisignKV{}
\global\long\def\usignKV{-\,}

\global\long\def\signKV{\,+\,}
\global\long\def\isignKV{\,-\,}
\global\long\def\usignKV{+}
\global\long\def\uisignKV{-\,}

\global\long\def\isignKV{\,+\,}
\global\long\def\signKV{\,-\,}
\global\long\def\uisignKV{}
\global\long\def\usignKV{-\,}

\global\long\def\signKV{\,+\,}
\global\long\def\isignKV{\,-\,}
\global\long\def\usignKV{}
\global\long\def\uisignKV{-\,}

\global\long\def\hrD{\hrColor D}
\global\long\def\hrQ{\hrColor Q}
\global\long\def\hrn{\hrColor n}
\global\long\def\hrDn{\hrColor{Dn}}
\global\long\def\hrx{\hrColor x}

\global\long\def\hrV{\hrColor V}
\global\long\def\hrU{\hrColor U}
\global\long\def\hrVbar{\hrColor{\bar{V}}}
\global\long\def\hrWbar{\hrColor{\bar{W}}}
\global\long\def\hrSV{\hrColor S}
\global\long\def\hrUtilde{\hrColor{\tilde{U}}}
\global\long\def\hrVubar{\hrColor{\underbar{V}}}

\global\long\def\hrD{\mathsfit{D}{\mkern1mu}}

\global\long\def\CN{\gSector{\mathcal{C}}}
 \global\long\def\CNdot{\dot{\gSector{\mathcal{C}}}}
 \global\long\def\gCvec{\gSector{\mathcal{C}}}
\global\long\def\CNsm#1{\CN[#1]}

\global\long\def\CL{\fSector{\widetilde{\fSector{\mathcal{C}}}}}
 \global\long\def\CLdot{\dot{\CL}}
 \global\long\def\fCvec{\fSector{\widetilde{\fSector{\mathcal{C}}}}}
\global\long\def\CLsm#1{\CL[#1]}

\global\long\def\Ctwo{\mSector{\mathcal{C}_{2}}}
 \global\long\def\Ctwodot{\dot{\mSector{\mathcal{C}}}_{\mSector 2}}
\global\long\def\Ctwosm#1{\Ctwo[#1]}

\global\long\def\gfCvec{\mSector{\mathcal{R}}}

\global\long\def\fsm{\xi}
\global\long\def\ksm{\eta}

\global\long\def\Cbim{\mSector{\mathcal{C}_{\mathrm{b}}}}

\global\long\def\gW{\gSector{W_{g}}}
 \global\long\def\fW{\fSector{W_{f}}}

\global\long\def\gCW{\gSector{\Omega_{g}}}
 \global\long\def\fCW{\fSector{\Omega_{f}}}

\global\long\def\fWA{A}
\global\long\def\fWB{B}
\global\long\def\fWC{C}
\global\long\def\fWD{D}

\fi

%% file: sec-10.tex
\section{Introduction}

In comparison to general relativity (GR), the phenomenology of the
Hassan-Rosen (HR) bimetric theory \cite{Hassan:2011zd,Hassan:2011ea,Hassan:2017ugh,Hassan:2018mbl}
is at an early stage, and the full extent of its physical features
is still unknown. In GR, numerical relativity plays an important role
in theoretical astrophysics to clarify structure formation and growth,
or formation processes of black holes. Only by means of numerical
simulation and experiments, it is possible to get a theoretical understanding
of such phenomena occurring in nature \cite{Shibata:2015nr,Baumgarte:2010nr,Gourgoulhon:2012trip,Alcubierre:2012intro,Bona:2009el}.
In particular, numerical relativity is essential in analyzing strong-field
gravitational systems to interpret gravitational wave detections \cite{Abbott:2016blz}.
Similar remarks also hold for the HR theory. In order to perform \emph{bimetric}
numerical simulations (and discriminate bimetric from GR predictions),
many questions need to be addressed; for example \cite{Shibata:1999va}:
\begin{enumerate}[itemsep=0.5ex,parsep=0pt,leftmargin=3em,label=(\arabic*)]
\item Which initial data solve the bimetric constraints for gravitational
collapse?
\item Which formalism to use for stable bimetric numerical evolution?
\item What are appropriate gauge conditions?
\item What algorithms provide numerical stability?
\item What are suitable boundary conditions for numerical grids?
\item What is a good method for finding the apparent horizon?
\end{enumerate}
To obtain reliable numerical results, all of these issues have to
be resolved.

This work belongs to a series of papers in which we aim to investigate
the above issues. Earlier papers in this series establish the bimetric
equations in the standard 3+1 form \cite{Kocic:2018ddp}, and calculate
the ratio of lapse functions for the case of spherical symmetry \cite{Kocic:2019zdy}.
In the current paper, we present a method for solving the bimetric
constraints to determine the initial data for the gravitational collapse
of dust in the HR theory. Using a numerical code based on the equations
of motion in the standard 3+1 form, we then obtain a short term development
of the initial data. Obtaining a long term bimetric development is
the subject of ongoing work. In order to achieve a more stable numerical
evolution, \cite{Torsello:2019tgc} establishes the covariant Baumgarte\textendash Shapiro\textendash Shibata\textendash Nakamura
(cBSSN) formalism \cite{Shibata:1995we,Baumgarte:1998te,Brown:2005a,Brown:2009a,Brown:2012a}
for bimetric relativity. A class of gauge conditions specific to the
HR theory are treated in \cite{Torsello:meang}. 

This paper is structured as follows. The rest of this section summarizes
the results by examples; it also reviews the basic equations. Section
\ref{sec:id} poses and solves the bimetric constraint equations.
Section \ref{sec:evol} highlights the properties of the evolution
of the initial data. The paper ends with a brief discussion and outlook.

\subsection{Summary of results}

We consider the case where the two metric sectors share the same spherical
symmetry \cite{Torsello:2017ouh}. In GR, the kinematical and dynamical
parts of a metric field can be separated using the 3+1 formalism \cite{Arnowitt:1962hi,York:1979aa}.
The same procedure can be applied to bimetric theory using a suitable
parametrization based on the geometric mean metric \cite{Kocic:2018ddp}.
In this context, the general form of the two metrics in spherical
polar coordinates reads,\bSe\label{eq:ssym-gf}
\begin{align}
\gMet & =-\gAlpha^{2}\dd t^{2}+\gEA^{2}(\dd r+\gBeta\,\dd t)^{2}+\gEB^{2}(\dd\theta^{2}+\sin^{2}\theta\,\dd\phi^{2}),\\
\fMet & =-\fAlpha^{2}\dd t^{2}+\fEA^{2}(\dd r+\fBeta\,\dd t)^{2}+\fEB^{2}(\dd\theta^{2}+\sin^{2}\theta\,\dd\phi^{2}),
\end{align}
\eSe where $\gAlpha$ and $\fAlpha$ denote the lapse functions,
$\gBeta$ and $\fBeta$ are the radial components of the shift vectors,
and $(\gEA,\gEB,\fEA,\fEB)$ denote the nontrivial components of the
spatial vielbeins. In addition, we also evolve the nonzero components
of the extrinsic curvature,
\begin{equation}
\gK_{1}=\tud{\gK}rr,\quad\gK_{2}=\tud{\gK}{\theta}{\theta}=\tud{\gK}{\phi}{\phi},\quad\fK_{1}=\tud{\fK}rr,\quad\fK_{2}=\tud{\fK}{\theta}{\theta}=\tud{\fK}{\phi}{\phi}.\label{eq:ssym-Ks}
\end{equation}
The radial components of the shift vectors are conveniently redefined
in terms of the mean shift $\hShift$ and the separation parameter
$\sLp$ \cite{Kocic:2018ddp},
\begin{equation}
\gBeta\coloneqq\hShift+\gAlpha\gEA^{-1}\sLv,\quad\fBeta\coloneqq\hShift-\fAlpha\fEA^{-1}\sLv,\quad\sLv\coloneqq\sLp\sLtinv,\quad\sLt\coloneqq(1+\sLp^{2})^{1/2}.\label{eq:ssym-shifts}
\end{equation}
In this parametrization, the lapses $\gLapse$, $\fLapse$, and the
mean shift $\hShift$ do not appear in the constraint equations. All
the variables in (\ref{eq:ssym-gf})\textendash (\ref{eq:ssym-shifts})
are functions of $(t,r)$. We let matter (if any) be coupled to only
one of the metric sectors, here $\gMet$. 

We start from the development of existing GR initial data for $\gMet$,
hereinafter called the\emph{ reference GR solution}. At this stage
we decouple $\fMet$; the bimetric theory is reduced to two copies
of GR where we only consider the evolution in the $\gMet$-sector
containing the collapsing matter (letting $\fMet$ evolve in parallel,
for instance, a Minkowski solution). The bimetric interaction will
be engaged in a second stage, treating the bimetric collapse after
knowing the behavior of the reference GR solution.

\begin{figure}[H]
\noindent \begin{centering}
\includegraphics{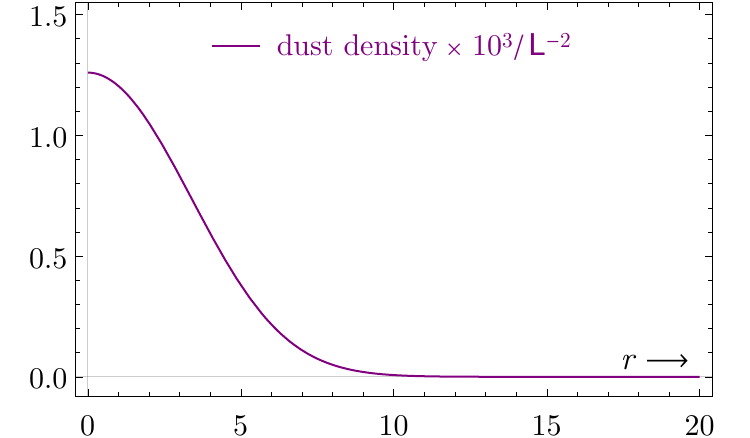}\includegraphics{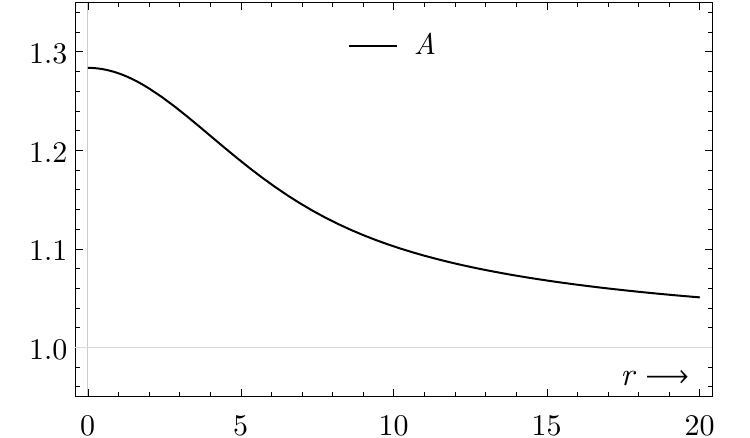}\vspace{-1ex}
\par\end{centering}
\caption{\label{fig1:gr}Initial data for the reference GR solution where the
density profile of dust is Gaussian at $t=0$. The radial coordinate
is in units of the black hole mass (formed after the collapse).}
\end{figure}

\clearpage\noindent As a reference GR solution, we consider a spherical
dust cloud with arbitrary density profile at $t=0$. The initial data
can be constructed using the time symmetric initial condition where
the extrinsic curvature is vanishing, assuming the conformally flat
spatial metric. The initial values for the density with a Gaussian
profile and the corresponding metric field $A$ are shown in figure
\ref{fig1:gr}. The development of this initial data forms a black
hole \cite{Nakamura:1980}.

\begin{figure}[H]
\noindent \begin{centering}
\hspace{-0.3em}\includegraphics{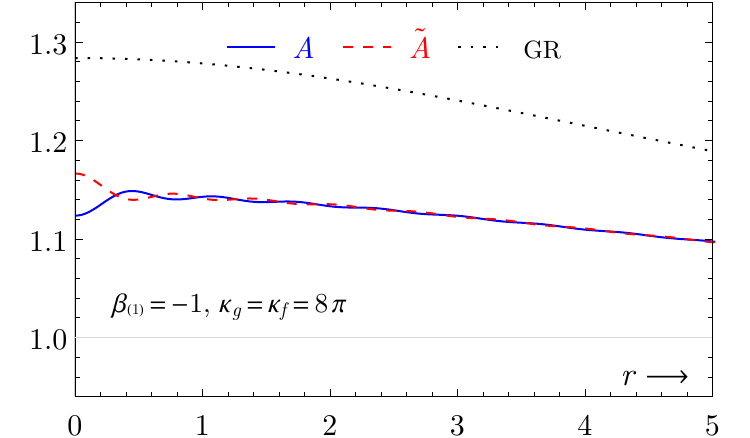}\includegraphics{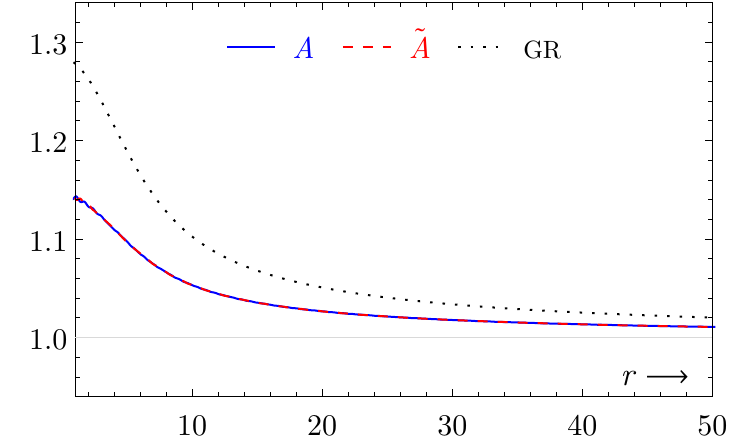}\vspace{-1ex}
\par\end{centering}
\caption{\label{fig2:bim-ex}An example of bimetric initial data obtained by
deforming the reference GR solution in figure \ref{fig1:gr}. The
bimetric fields $\protect\gEA$ and $\protect\fEA$ are interwoven
and oscillate along the radial coordinate.}
\end{figure}

After constructing the reference GR solution, we engage the two sectors
through the bimetric interaction and solve the constraint equations
for the metric fields, keeping the same matter profile coupled to
$\gMet$. This results in a system of coupled differential equations
whose solution depends on the parameters of the HR theory. The construction
of the bimetric initial data is the topic of section \ref{sec:id}.
Typical initial values for the metric fields $\gEA$ and $\fEA$ which
are far from the GR limit are shown in figure \ref{fig2:bim-ex}.
These fields carry physical content for the observers in the respective
metric sector; they roughly correspond to gravitational potentials
in the Newtonian limit.

\begin{figure}[H]
\noindent \begin{centering}
\hspace{5em}\includegraphics[scale=0.75]{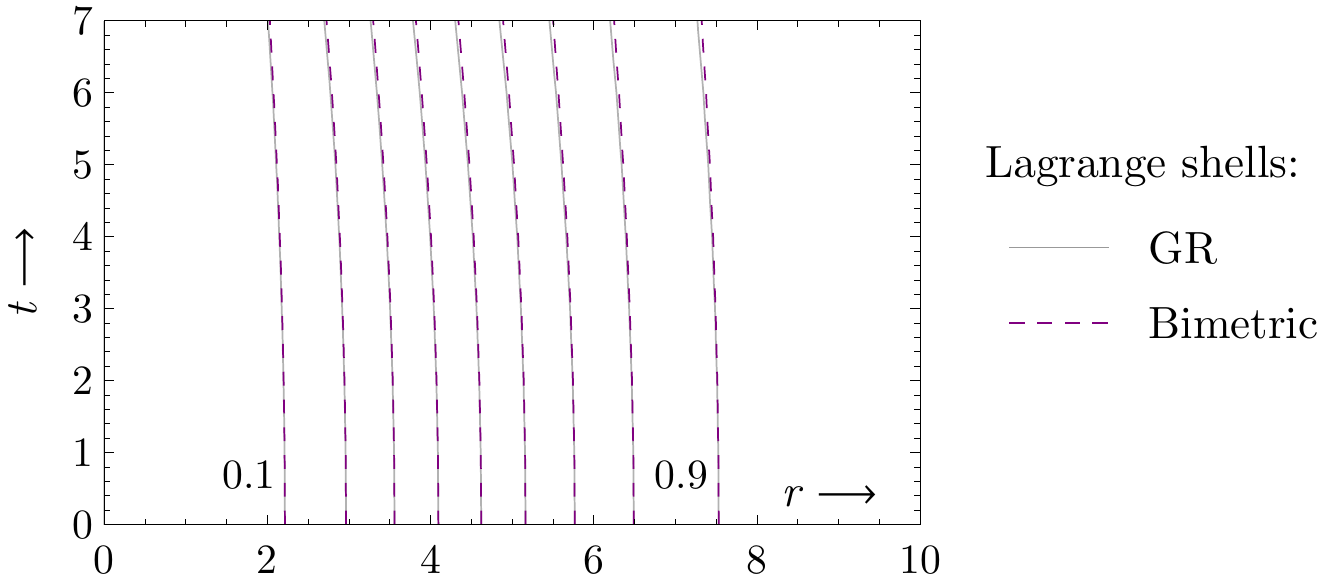}\vspace{-2ex}
\par\end{centering}
\caption{\label{fig3:gr-bim}Lagrange shells for the reference GR and the bimetric
initial data which are close to GR (the shells are worldlines of Lagrangian
matter tracers labeled by the fixed interior rest-mass fraction).
The collapse is in the strong-field regime, where almost all of the
dust is initially inside $r\approx8$. In GR, the apparent horizon
of the formed black hole is at $r\approx3$ after $t\approx40$ (figure
\ref{fig5:ref-gr}). The bimetric solution follows GR, showing no
instabilities during the initial phase (see subsection \ref{3.4}).}
\end{figure}

Finally, as described in section \ref{sec:evol}, we numerically evolve
the initial data. The simulations are not long enough to shed light
on the end point of gravitational collapse. Nevertheless, the bimetric
collapse of the dust cloud stays very close to the reference GR solution,
as illustrated in figure \ref{fig3:gr-bim}. More details about the
gravitational collapse in the GR limit are given in subsection \ref{3.4}
(see also figures \ref{fig5:ref-gr} and \ref{fig17:sim-collapse}).

\begin{figure}[H]
\noindent \begin{centering}
\hspace{-1mm}\includegraphics{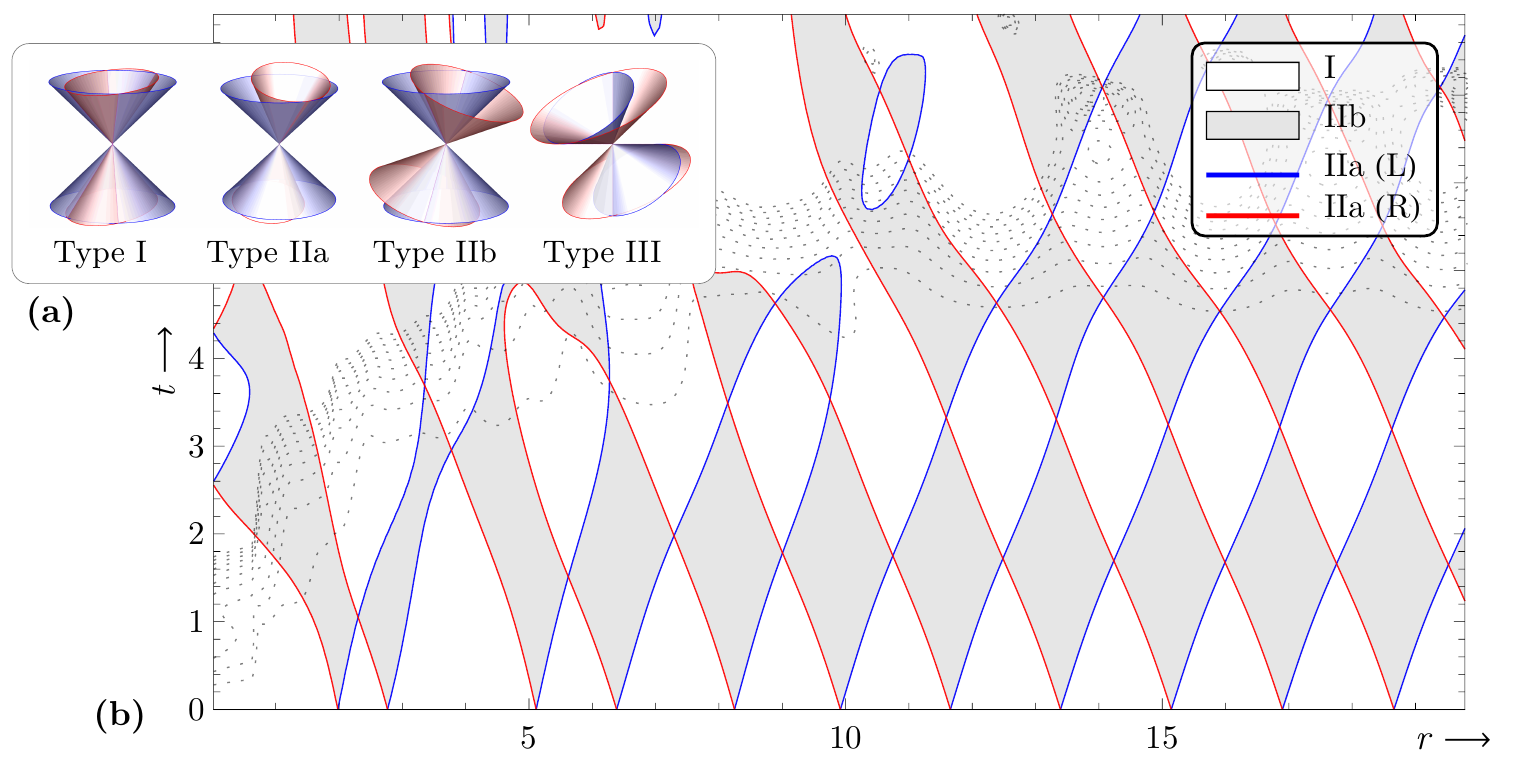}\vspace{-3ex}
\par\end{centering}
\caption{\label{fig4:cd-ex}Causal diagram for the evolution of bimetric initial
data. The white regions are Type I (bidiagonal), the shaded Type IIb,
and the edges Type IIa. Type III does not appear in spherical symmetry
because of the dimensional reduction. The dotted lines are surface
levels of the $l^{2}$-norm of the constraint violations (with separation
$10^{-3}$ and no violations at $t=0$); the result is trustworthy
for $t<5$ at large radii. The plot displays the oscillations of the
metric fields in space and time.}
\end{figure}

To highlight the bimetric features in the physical content, we employ
causal diagrams that display metric configurations according to the
method given in \cite{Kocic:cdg}. From these diagrams one can read-off
the oscillation frequencies of the metric fields in space and time.
The causal diagrams are based on the fact that the possible configurations
of two metrics can be visualized in terms of the null cone intersections
\cite{Hassan:2017ugh}, as shown in figure~\ref{fig4:cd-ex}(a).
Coordinate transformations only deform the null cones, keeping the
nature of their intersections. This gives an invariant picture of
the bimetric spacetime. The causal diagram for the evolution of the
initial data from figure~\ref{fig2:bim-ex} is plotted in figure~\ref{fig4:cd-ex}(b),
showing the leading frequencies in the dynamics of the metrics fields.\vspace{-0.5ex}

Figure \ref{fig4:cd-ex} also displays how the metric configurations
are interwoven in space and time. Note that the metrics have been
constructed to be simultaneously diagonalizable at the initial hypersufrace.
Such ``bidiagonal slices'' may locally reoccur during the evolution
(horizontally slicing the white rhomboid-like patches in some regions
in figure~\ref{fig4:cd-ex}). Notwithstanding, the two metrics are
in general not simultaneously diagonalizable, and the solution lacks
a timelike Killing vector field (so it cannot be made stationary by
any coordinate transformation). Finally, the dotted lines in figure~\ref{fig4:cd-ex}
indicate the regions where the constraints get more violated. These
are the artifacts of the numerical simulation and indicate a departure
from the bimetric solution. More elaborate comments on the evolution
and numerical simulations are found in section~\ref{sec:evol}.

\subsection{Basic equations}

We consider the metrics $\gMet$ and $\fMet$ coupled through the
ghost-free bimetric potential \cite{deRham:2010ik,deRham:2010kj,Hassan:2011vm},%
\bgroup%
\setlength\abovedisplayskip{0.7ex}%
\setlength\belowdisplayskip{0.7ex}%
\begin{equation}
V(S)\coloneqq\usignV\betaSumL\,e_{n}(S),\qquad S\coloneqq(\gMet^{-1}\fMet)^{1/2}.\label{eq:V}
\end{equation}
\egroup Here, $(\gMet^{-1}\fMet)^{1/2}$ denotes the principal square
root of the (1,1) tensor field  $\gMet^{\mu\rho}\fMet_{\rho\nu}$,
and $e_{n}(S)$ are the elementary symmetric polynomials (the principal
scalar invariants of $S$). The potential is parametrized by dimensionless
real constants $\beta_{(n)}$ with an overall length scale $\betaScale$
in the geometrized units $c=G=1$. The particular form of the potential
is dictated by the necessary condition for the absence of ghosts \cite{Creminelli:2005qk},
where the dynamics of each metric is given by a separate Einstein-Hilbert
term in the action \cite{Hassan:2011zd}. The principal branch of
the square root ensures an unambiguous definition of the theory \cite{Hassan:2017ugh}.

The resulting bimetric field equations are (here stated in the operator
form),\bSe\label{eq:bim-eom}
\begin{alignat}{2}
\gEinst & =\gKappa(\gVse+\gTse), & \qquad\gVse & \coloneqq\usignV\betaSum\,Y_{n}(S),\\[1ex]
\fEinst & =\fKappa(\fVse+\fTse), & \qquad\fVse & \coloneqq\usignV\betaSum\,Y_{d-n}(S^{-1}),
\end{alignat}
\eSe where $\gEinst$ and $\fEinst$ are the Einstein tensors of
the two metrics, $\gKappa$ and $\fKappa$ are Einstein's gravitational
constants, $\gTse$ and $\fTse$ are the stress\textendash energy
tensors of the matter fields each minimally coupled to a different
metric sector, and $\gVse$ and $\fVse$ are the contributions of
the bimetric potential (\ref{eq:V}), also called the bimetric stress\textendash energy
tensors. The functions $Y_{n}(S)$ in (\ref{eq:bim-eom}) encapsulates
the variation of the bimetric potential with respect to the metrics,%
\bgroup%
\setlength\abovedisplayskip{0.7ex}%
\setlength\belowdisplayskip{0.7ex}%
\begin{gather}
Y_{n}(S)\coloneqq\sum_{k=0}^{n}(-1)^{n+k}e_{k}(S)\,S^{n-k}=\frac{\partial e_{n+1}(S)}{\partial S^{\tr}},\qquad Y_{n<0}(S)=Y_{n\ge4}(S)\equiv0.\label{eq:Yn-def}
\end{gather}
\egroup The bimetric stress\textendash energy tensors satisfy the
following identities \cite{Hassan:2014vja,Damour:2002ws},\bSe\label{eq:bim-ids}
\begin{align}
\sqrt{-\gMet}\,\tud{\gVse}{\mu}{\nu}+\sqrt{-\fMet}\,\tud{\fVse}{\mu}{\nu}-\sqrt{-\gMet}\,V\,\tud{\delta}{\mu}{\nu} & =0,\label{eq:id-alg2}\\[1ex]
\sqrt{-\gMet}\,\gCD_{\mu}\gVse{}^{\mu}{}_{\nu}+\sqrt{-\fMet}\,\fCD_{\mu}\fVse{}^{\mu}{}_{\nu} & =0,\label{eq:id-damour}
\end{align}
\eSe where $\gCD_{\mu}$ and $\fCD_{\mu}$ are the covariant derivatives
compatible with $\gMet$ and $\fMet$, respectively. Assuming that
the matter conservation laws hold, $\gCD_{\mu}\tud{\gTse}{\mu}{\nu}=0$
and $\fCD_{\mu}\tud{\fTse}{\mu}{\nu}=0$, the field equations (\ref{eq:bim-eom})
imply the bimetric conservation law,
\begin{equation}
\gCD_{\mu}\tud{\gVse}{\mu}{\nu}=0,\qquad\fCD_{\mu}\tud{\fVse}{\mu}{\nu}=0.\label{eq:bianchi}
\end{equation}
The two equations in (\ref{eq:bianchi}) are not independent according
to the differential identity (\ref{eq:id-damour}).

\paragraph*{The 3+1 decomposition.}

The structure of the bimetric field equations is equivalent to having
two copies of GR with the additional stress\textendash energy contributions
$\gVse$ and $\fVse$ coupled through (\ref{eq:id-alg2}); hence,
their 3+1 split is straightforward, provided that one knows the projections
of $\gVse$ and $\fVse$ \cite{Kocic:2018ddp}. The 3+1 decomposition
of (\ref{eq:bim-eom}) results in two sets of the evolution and constraint
equations formally analog to those in GR \cite{Shibata:2015nr,Baumgarte:2010nr,Gourgoulhon:2012trip,Alcubierre:2012intro,Bona:2009el}. 

The constraint equations do not depend on the lapse functions and
one shift vector because of the particular form of the potential (\ref{eq:V}).
The projection of the bimetric conservation law (\ref{eq:bianchi})
gives one additional constraint \cite{Kocic:2018ddp}, equivalent
to the additional constraint obtained from the Hamiltonian analysis
\cite{Hassan:2018mbl} (the so-called secondary constraint). Moreover,
as shown in \cite{Alexandrov:2012yv,Hassan:2018mbl}, the preservation
of this additional constraint relates the two lapses through a ratio,
calculated for the case of spherical symmetry in \cite{Kocic:2019zdy}. 

Since our focus is not on the evolution equations, their standard
3+1 form is given in appendix \ref{app:ev-eqs}. The constraint equations
are treated in the following section in more detail. 

%% file: sec-20.tex
\section{Initial data}

\label{sec:id}

The initial values for evolution are subject to certain constraints.
In spherical symmetry, the dynamical variables are $(\gEA,\gEB,\gK_{1},\gK_{2})$
and $(\fEA,\fEB,\fK_{1},\fK_{2})$. Also, the separation parameter
$\sLp$ defines the relative shift between the two metrics (characterizing
the relative `off-diagonality' of the metrics). The two lapse functions
$\gLapse$, $\fLapse$, and the mean shift $\hShift$ are kinematical
variables and do not appear in the constraint equations. Hence, we
need to specify the following metric functions of $r$ at $t=0$:
$\gEA$, $\gEB$, $\gK_{1}$, $\gK_{2}$, $\fEA$, $\fEB$, $\fK_{1}$,
$\fK_{2}$, and $\sLp$. These functions are constrained by five equations,
which leaves a lot of freedom in choosing their initial values. In
the rest of this section, we state the constraint equations, solve
for a reference GR solution, and construct the related bimetric initial
data.

\subsection{Constraint equations}

To write down the constraint equations, we need the 3+1 decomposition
of the stress\textendash en\-er\-gy tensors. Let $\grho,\gjota_{i},\tud{\gJota}ij,\frho,\fjota_{i},$
and $\tud{\fJota}ij$ denote the normal, tangential and spatial projections
of the \emph{total} stress\textendash energy tensors $V+T$ in the
respective sector. These projections sum up the contribution coming
from the bimetric potential, denoted by the upper label ``b'', and
the matter contribution, denoted by the upper label ``m''; for example,
$\grho=\grho^{\gblab}+\grho^{\gmlab}$. 

\paragraph*{Bimetric sources.}

The nonzero components of the projections of $\gVse$ are \cite{Kocic:2018ddp},\bSe\label{eq:ssym-g-se}\vspace{-0.5ex}
\begin{align}
\grho^{\gblab} & =-\bigg[\left\langle \sER\right\rangle _{0}^{2}+\sLt\frac{\fEA}{\gEA}\left\langle \sER\right\rangle _{1}^{2}\bigg],\qquad\gjota_{r}^{\gblab}=-\sLp\fEA\left\langle \sER\right\rangle _{1}^{2},\\
\gJota_{1}^{\gblab} & =\left\langle \sER\right\rangle _{0}^{2}+\bigg[\frac{1}{\sLt}\bigg(\frac{\fAlpha}{\gAlpha}+\frac{\fEA}{\gEA}\bigg)-\sLt\frac{\fEA}{\gEA}\bigg]\left\langle \sER\right\rangle _{1}^{2},\\
\gJota_{2}^{\gblab} & =\left\langle \sER\right\rangle _{0}^{1}+\frac{\fAlpha\fEA}{\gAlpha\gEA}\left\langle \sER\right\rangle _{1}^{2}+\frac{1}{\sLt}\bigg(\frac{\fAlpha}{\gAlpha}+\frac{\fEA}{\gEA}\bigg)\left\langle \sER\right\rangle _{1}^{1},
\end{align}
\eSe where $\gJota_{1}\coloneqq\tud{\gJota}rr$, $\gJota_{2}\coloneqq\tud{\gJota}{\theta}{\theta}=\tud{\gJota}{\phi}{\phi}$,
$\gJota=\gJota_{1}+2\gJota_{2}$.\newpage

\noindent Similarly for $\fVse$ we have,\bSe \label{eq:ssym-f-se}\vspace{-0.5ex}
\begin{align}
\frho^{\fblab} & =-\bigg[\left\langle \sER\right\rangle _{2}^{2}+\sLt\frac{\gEA}{\fEA}\left\langle \sER\right\rangle _{1}^{2}\bigg]\frac{1}{\sER^{2}},\qquad\fjota_{r}^{\fblab}=\sLp\gEA\left\langle \sER\right\rangle _{1}^{2}\frac{1}{\sER^{2}},\\
\fJota_{1}^{\fblab} & =\bigg\{\left\langle \sER\right\rangle _{2}^{2}+\bigg[\frac{1}{\sLt}\bigg(\frac{\gAlpha}{\fAlpha}+\frac{\gEA}{\fEA}\bigg)-\sLt\frac{\gEA}{\fEA}\bigg]\left\langle \sER\right\rangle _{1}^{2}\bigg\}\frac{1}{\sER^{2}},\\
\fJota_{2}^{\fblab} & =\bigg\{\left\langle \sER\right\rangle _{3}^{1}+\frac{\gAlpha\gEA}{\fAlpha\fEA}\left\langle \sER\right\rangle _{1}^{1}+\frac{1}{\sLt}\bigg(\frac{\gAlpha}{\fAlpha}+\frac{\gEA}{\fEA}\bigg)\left\langle \sER\right\rangle _{2}^{1}\bigg\}\frac{1}{\sER},
\end{align}
\eSe where $\fJota_{1}\coloneqq\tud{\fJota}rr$, $\fJota_{2}\coloneqq\tud{\fJota}{\theta}{\theta}=\tud{\fJota}{\phi}{\phi}$,
and $\fJota=\fJota_{1}+2\fJota_{2}$. To simplify equations, we have
defined $\sER\coloneqq\fEB/\gEB$ and,
\begin{equation}
\left\langle \sER\right\rangle _{k}^{n}\coloneqq\usignV\betaScale\sum_{i=0}^{n}\mbinom ni\beta_{(i+k)}\sER^{i},\quad\ \left\langle \sER\right\rangle _{k}^{n}=\left\langle \sER\right\rangle _{k}^{n-1}+\sER\left\langle \sER\right\rangle _{k+1}^{n-1},\ \left\langle \sER\right\rangle _{k}^{0}=\usignV\betaScale\beta_{(k)}.
\end{equation}
The function $\left\langle \cdot\right\rangle _{k}^{n}$ encapsulates
the $\beta_{(k)}$-parameters that do not appear elsewhere. The tangential
components satisfy the following identity coming from (\ref{eq:id-alg2}),
\begin{equation}
\sqrt{\gSp}\,\gjota_{r}^{\gblab}+\sqrt{\fSp}\,\fjota_{r}^{\fblab}=0,\qquad\text{where }\,\sqrt{\gSp}\coloneqq\gEA\gEB^{2}\,\text{ and }\,\sqrt{\fSp}\coloneqq\fEA\fEB^{2}.\label{eq:id-alg-ss}
\end{equation}

\paragraph*{Matter sources.}

We assume that matter is present only in the $\gMet$-sector and let
$\gSector T\coloneqq\gTse$. In particular, we consider a perfect
fluid with the stress\textendash energy tensor \cite{Baumgarte:2010nr,Rezzolla:2013rel},
\begin{equation}
\gSector T_{\mu\nu}=\pfrho\pfh\,\pfu_{\mu}\pfu_{\nu}+\pfP\gMet_{\mu\nu},\qquad\pfh\coloneqq1+\pfeps+\pfP/\pfrho,\qquad\gMet_{\mu\nu}\pfu^{\mu}\pfu^{\nu}=-1,
\end{equation}
where $\pfrho$ is the rest-mass density, $\pfh$ is the specific
enthalpy, $\pfeps$ is the specific energy, $\pfP$ is the pressure,
and $\pfu^{\mu}$ is the four\textendash velocity of the fluid. The
general relativistic hydrodynamic equations consist of the conservation
law for $\gSector T_{\mu\nu}$, the conservation law of the baryon
number, and the equation of state for the fluid, respectively given
by,
\begin{equation}
\gCD_{\mu}\gSector T^{\mu\nu}=0,\qquad\gCD_{\mu}(\pfrho\pfu^{\mu})=0,\qquad\pfP(\pfrho,\pfeps)=0.\label{eq:pf-eqs}
\end{equation}
Following the 3+1 ``Valencia'' formulation \cite{Banyuls:1997zz,Rezzolla:2013rel},
we rewrite (\ref{eq:pf-eqs}) in the first order flux\textendash conservative
form. In spherical coordinates, the resulting equations are expressed
in terms of the following conserved\emph{ }variables (here densitizied),
\begin{equation}
\pfD\coloneqq\sqrt{\gSp}\pfrho\pfW,\qquad\pfS_{r}\coloneqq\sqrt{\gSp}\pfrho\pfh\pfW^{2}\gEA^{2}\pfv,\qquad\pftau\coloneqq\sqrt{\gSp}\left(\pfrho\pfh\pfW^{2}-\pfP\right)-\pfD,
\end{equation}
where $\sqrt{\gSp}$ is defined in (\ref{eq:id-alg-ss}), $\pfv$
is the radial component of the Eulerian three-velocity of the fluid,
and $w$ is the corresponding Lorentz factor,
\begin{equation}
\pfv\coloneqq(\pfu^{r}/\pfu^{t}+\gBeta)/\gLapse,\qquad\pfW\coloneqq\gLapse\pfu^{t}=1/\sqrt{1-\gEA^{2}\pfv^{2}}.
\end{equation}
The flux\textendash conservative evolution equations for $(\pfD,\pfS_{r},\pftau)$
are relegated to the ancillary file. 

For a pressureless fluid (dust), we have $\pfP=0$, $\pfeps=0$, $\pfh=1$,
and the conversion from the conserved to the primitive\emph{ }variables
reads,
\begin{equation}
\pfrho=\frac{\pfD}{\pfW\sqrt{\gSp}},\quad\pfv=\frac{\pfS_{r}\gEA^{-1}}{\pftau+\pfD},\quad\pfW=\frac{\pftau+\pfD}{\sqrt{(\pftau+\pfD)^{2}-\pfS_{r}^{2}\gEA^{-2}}}=\sqrt{1+\frac{\pfS_{r}^{2}\gEA^{-2}}{\pfD^{2}}}.
\end{equation}
Finally, the calculation in \cite{Rezzolla:2013rel} yields the components
of the matter stress\textendash en\-er\-gy tensor,\bSe\label{eq:pf-adm}
\begin{align}
\grho^{\gmlab} & =\pfrho\pfW^{2}=(\pftau+\pfD)/\sqrt{\gSp},\\
\gjota_{r}^{\gmlab} & =\pfrho\pfW^{2}\gEA^{2}\pfv=\pfS_{r}/\sqrt{\gSp},\\
\gJota_{1}^{\gmlab} & =\pfv\pfS_{r},\qquad\gJota_{2}^{\gmlab}=0.
\end{align}
\eSe

\paragraph*{Bimetric constraints.}

The scalar and vector constraint equations are obtained as the normal
and tangential projections of the bimetric field equations (\ref{eq:bim-eom})
on the initial hypersurface. The scalar (or Hamiltonian) constraint
equations are,\bSe\label{eq:scC}
\begin{alignat}{2}
C_{1} & \coloneqq(2\gK_{1}+\gK_{2})\gK_{2}+\frac{1}{\gEA^{2}}\bigg(\frac{\gEA^{2}}{\gEB^{2}}+2\frac{\partial_{r}\gEA}{\gEA}\frac{\partial_{r}\gEB}{\gEB}-\frac{(\partial_{r}\gEB)^{2}}{\gEB^{2}}-2\frac{\partial_{r}^{2}\gEB}{\gEB}\bigg) & -\,\gKappa(\grho^{\gblab}+\grho^{\gmlab}) & =0,\label{eq:scCg}\\
C_{2} & \coloneqq(2\fK_{1}+\fK_{2})\fK_{2}+\frac{1}{\fEA^{2}}\bigg(\frac{\fEA^{2}}{\fEB^{2}}+2\frac{\partial_{r}\fEA}{\fEA}\frac{\partial_{r}\fEB}{\fEB}-\frac{(\partial_{r}\fEB)^{2}}{\fEB^{2}}-2\frac{\partial_{r}^{2}\fEB}{\fEB}\bigg) & -\,\fKappa(\frho^{\fblab}+\frho^{\fmlab}) & =0.\label{eq:scCf}
\end{alignat}
The vector (or momentum) constraint equations are, 
\begin{alignat}{2}
C_{3} & \coloneqq(\gK_{1}-\gK_{2})\frac{\partial_{r}\gEB}{\gEB}-\partial_{r}\gK_{2} & \,\signK\frac{1}{2}\gKappa(\gjota_{r}^{\gblab}+\gjota_{r}^{\gmlab}) & =0,\label{eq:vecCg}\\
C_{4} & \coloneqq(\fK_{1}-\fK_{2})\frac{\partial_{r}\fEB}{\fEB}-\partial_{r}\fK_{2} & \,\signK\frac{1}{2}\fKappa(\fjota_{r}^{\fblab}+\fjota_{r}^{\fmlab}) & =0.\label{eq:vecCf}
\end{alignat}
\eSe The last two equations can be combined using the identity (\ref{eq:id-alg-ss}),
\begin{align}
\fKappa\fEA\gEB\big(\gK_{1}\partial_{r}\gEB-\gK_{2}\partial_{r}\gEB-\gEB\partial_{r}\gK_{2}-\mfrac 12\gKappa\gjota_{r}^{\gmlab}\big)\qquad\qquad\quad\nonumber \\
+\,\gKappa\gEA\fEB\big(\fK_{1}\partial_{r}\fEB-\fK_{2}\partial_{r}\fEB-\fEB\partial_{r}\fK_{2}-\mfrac 12\fKappa\fjota_{r}^{\fmlab}\big) & =0.
\end{align}
The separation parameter $\sLp$ can be determined either from (\ref{eq:vecCg})
or (\ref{eq:vecCf}). 

The bimetric scalar and vector constraints (\ref{eq:scC}) are the
same as in GR with the addition of the bimetric sources; what is specific
to the HR theory is the additional constraint obtained from the projection
of the bimetric conservation law,
\begin{align}
C_{\mathrm{bim}} & \coloneqq\fEA\Big(\fK_{1}\left\langle \sER\right\rangle _{1}^{2}+2\fK_{2}\sER\left\langle \sER\right\rangle _{2}^{1}\Big)+2\gEA\fK_{2}\sLt\sER\left\langle \sER\right\rangle _{1}^{1}-\gEA\Big(\gK_{1}\left\langle \sER\right\rangle _{1}^{2}+2\gK_{2}\left\langle \sER\right\rangle _{1}^{1}\Big)\nonumber \\
 & \qquad-\,2\fEA\gK_{2}\sLt\left\langle \sER\right\rangle _{2}^{1}\isignK2\sLp\bigg(\left\langle \sER\right\rangle _{1}^{1}\frac{\gEA}{\fEA}\frac{\partial_{r}\fEB}{\gEB}+\left\langle \sER\right\rangle _{2}^{1}\frac{\fEA}{\gEA}\frac{\partial_{r}\gEB}{\gEB}\bigg)\isignK\sLtinv\left\langle \sER\right\rangle _{1}^{2}\partial_{r}\sLp=0.\label{eq:ssym-cc2}
\end{align}
As an example of the bimetric initial data with no matter sources,
consider the initial values $\sLp=0$, $\gEA=\fEA=1$, $\gEB=\fEB=r$,
and $\gK_{1}=\gK_{2}=\fK_{1}=\fK_{2}=0$ where $\left\langle 1\right\rangle _{0}^{3}=\left\langle 1\right\rangle _{1}^{3}=0$.
These satisfy (\ref{eq:scC})\textendash (\ref{eq:ssym-cc2}) and
yield the bi-Minkowski solution.

%% file: sec-22.tex
\subsection{Reference GR solution}

As mentioned in the introduction, we start from a reference GR solution
with decoupled bimetric sectors; we set $\beta_{(n)}=0$ and let the
two sectors independently evolve in parallel. This implies $\left\langle \cdot\right\rangle _{k}^{n}\equiv0$,
hence (\ref{eq:ssym-cc2}) identically vanishes at all times. The
$\fMet$-sector is set up to develop the Minkowski solution from the
initial data $\fEA=1$, $\fEB=r$, and $\fK_{1}=\fK_{2}=0$. 

\begin{figure}[H]
\noindent \centering{}\includegraphics[width=150mm]{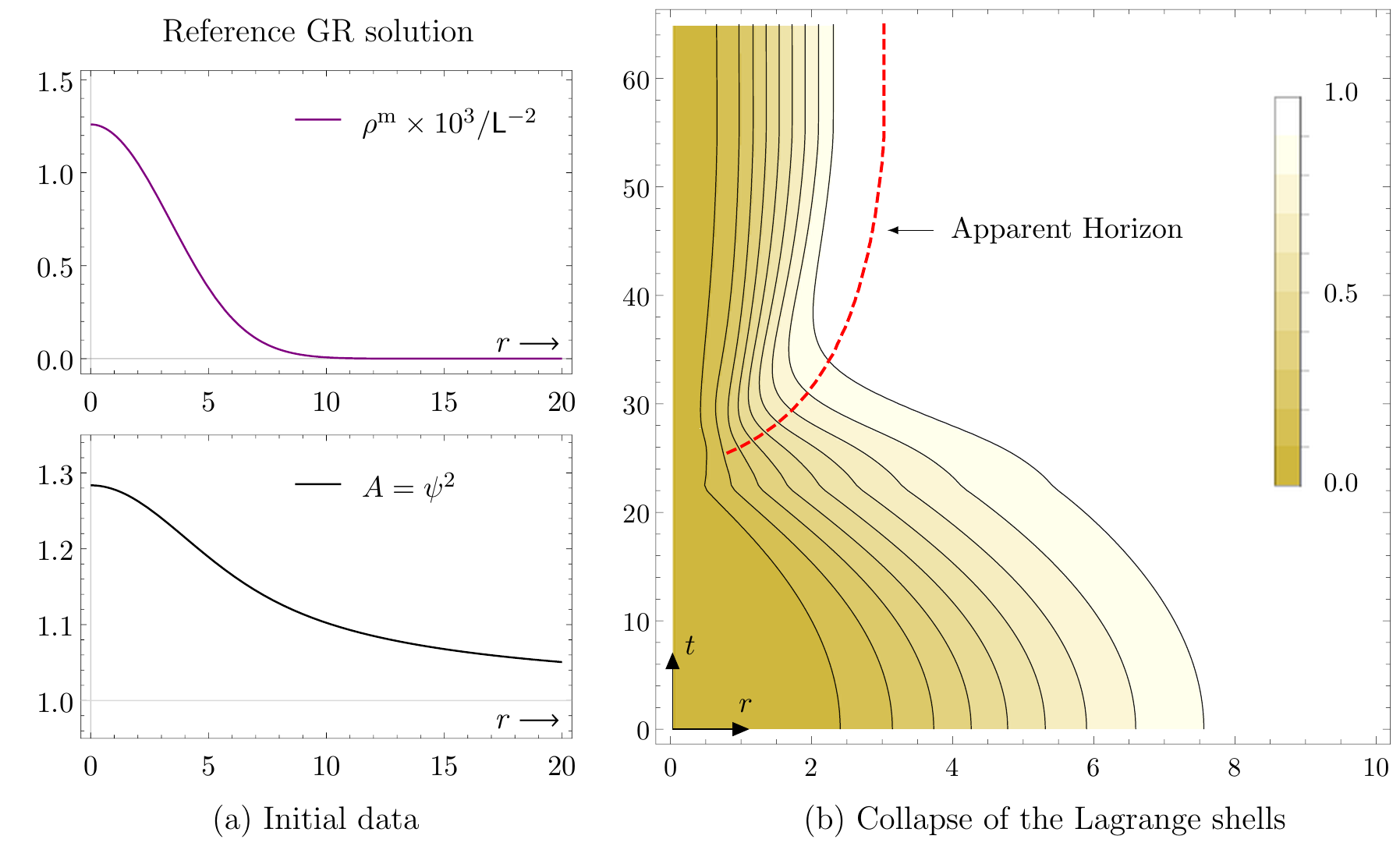} \vspace{-2ex}\caption{\label{fig5:ref-gr}(a) The initial data of the reference GR solution.
(b) The development of the initial data forms a black hole. The dashed
line shows the apparent horizon.}
\end{figure}

\clearpage\noindent The initial data for $\gMet$ are constructed
using the time symmetric initial condition,
\begin{equation}
\gK_{1}=\gK_{2}=0,\qquad\pftau=\pfv=0\ \Rightarrow\ \gjota_{r}^{\gmlab}=\pfS_{r}=0,
\end{equation}
also assuming a conformally flat spatial metric at $t=0$ with the
conformal factor $\cphi(r)$,
\begin{equation}
\gEA=\cphi^{2},\qquad\gEB=\cphi^{2}r,\qquad\dd\ell_{\gMet}^{2}=\cphi^{4}\left(\dd r^{2}+r^{2}\dd\theta^{2}+r^{2}\sin^{2}\theta\,\dd\phi^{2}\right).
\end{equation}
At this point all the bimetric constraints except (\ref{eq:scCg})
are satisfied. Hence, $\cphi$ and $\pfD$ must satisfy,
\begin{equation}
4\cphi\Delta\cphi+\gKappa\pfD=0,\qquad\Delta\psi\coloneqq\frac{1}{r^{2}}\frac{\partial}{\partial r}\left(r^{2}\frac{\partial}{\partial r}\psi\right),\label{eq:ideqgr}
\end{equation}
where $\grho^{\gmlab}=\pfD\cphi^{-6}$ and $\Delta$ denotes the spherical
Laplacian. Consider the initial density,
\begin{equation}
\pfD(r)=c_{0}^{3}\left[1+\frac{1}{2r}\op{erf}(c_{0}r\sqrt{\pi})\right]\ee^{-c_{0}^{2}r^{2}\pi},\quad\op{erf}(z)\coloneqq\frac{2}{\sqrt{\pi}}\int_{0}^{z}\ee^{-x^{2}}\dd x.
\end{equation}
Here, $c_{0}>0$ is a free parameter and $\op{erf}(z)$ is the integral
of the Gaussian distribution. This density is the same as in \cite{Nakamura:1980}
for $\gKappa=8\pi$. The solution to (\ref{eq:ideqgr}) becomes,
\begin{equation}
\cphi(r)=1+\frac{1}{2r}\op{erf}(c_{0}r\sqrt{\pi}),\qquad\lim_{r\,\to\,0}\cphi(r)=1+c_{0},\quad\lim_{r\,\to\,\infty}\cphi(r)=1.
\end{equation}
An example of the initial data for $c_{0}=(3\sqrt{2\pi})^{-1}\approx0.133$
is shown in figure \ref{fig5:ref-gr}; the evolution of these data
is treated in section \ref{sec:evol}, with the results shown in the
right panel of figure \ref{fig5:ref-gr}.\vspace{1ex}

%% file: sec-23.tex
\subsection{Bimetric polytrope}

\label{ssec:bimpolytrope}

We now engage the bimetric interaction, keeping the same initial density
of dust as in the reference GR solution. We again use the time symmetric
initial condition with the conformally flat spatial metrics in each
sectors,\bSe
\begin{alignat}{3}
\gEA & =\gcphi^{2}, & \qquad\gEB & =\gcphi^{2}r, & \qquad\dd\ell_{\gMet}^{2} & =\gcphi^{4}\left(\dd r^{2}+r^{2}\dd\theta^{2}+r^{2}\sin^{2}\theta\,\dd\phi^{2}\right),\\
\fEA & =\fcphi^{2}, & \qquad\fEB & =\fcphi^{2}r, & \qquad\dd\ell_{\fMet}^{2} & =\fcphi^{4}\left(\dd r^{2}+r^{2}\dd\theta^{2}+r^{2}\sin^{2}\theta\,\dd\phi^{2}\right).
\end{alignat}
\eSe We set $\sLp=0$ but allow for arbitrary $\beta_{(n)}$-parameters.
The metric configuration is therefore of Type I at the initial hypersurface
with simultaneously diagonal $\gMet$ and $\fMet$.

At the moment of time symmetry where $\gK_{1}=\gK_{2}=\fK_{1}=\fK_{2}=0$,
$\pfv=0$, and $\sLp=0$, the constraint equations (\ref{eq:vecCg}),
(\ref{eq:vecCf}), and (\ref{eq:ssym-cc2}) are automatically satisfied.
Thus, we only have to consider the scalar constraints (\ref{eq:scCg})
and (\ref{eq:scCf}); these form a system of generalized Lane\textendash Emden
equations,\bSe\label{eq:ideqbim}\vspace{0.5ex}
\begin{alignat}{2}
4\gcphi\Delta\gcphi & \,+\,\gKappa\betaScale\left(\betap 0\gcphi^{6}+3\betap 1\fcphi^{2}\gcphi^{4}+3\betap 2\fcphi^{4}\gcphi^{2}+\betap 3\fcphi^{6}\right) & \,+\,\gKappa\pfD & =0,\label{eq:ideqbim1}\\
4\fcphi\Delta\fcphi & \,+\,\fKappa\betaScale\left(\betap 1\gcphi^{6}+3\betap 2\fcphi^{2}\gcphi^{4}+3\betap 3\fcphi^{4}\gcphi^{2}+\betap 4\fcphi^{6}\right) &  & =0.
\end{alignat}
\eSe In astrophysics \cite{Chandra:1957intro}, the Lane\textendash Emden
equation represents  a self\textendash gravitating spherically symmetric
fluid with the polytropic equation of state $P\propto\rho^{1+1/n}$;
in that context, $\rho\propto\vartheta^{n}$ satisfies the Lane\textendash Emden
equation $\Delta\vartheta+\vartheta^{n}=0$. Now, interpreting the
conformal factors as gravitational potentials in the Newtonian limit,
the metric fields obey deformed polytropic equations of states (where
the polytropic index becomes definite for a specific $\beta_{(k)}$-model
in vacuum, $\pfD=0$). In other words, engaging the second metric
in an existing GR solution introduces a fluid with nontrivial features
in both sectors; this gives the name ``bimetric polytrope'' (referred
to as ``pure'' when $\pfD\equiv0$). The departure from GR is controlled
by the overall factors $\gKappa\betaScale$ and $\fKappa\betaScale$. 

Given $\pfD(r)$,  the system (\ref{eq:ideqbim}) should be solved
for $\gcphi(r)$ and $\fcphi(r)$. We choose asymptotic boundary conditions
where $\pfD\to0$, $\gcphi\to\mathrm{const}$, and $\fcphi^{2}/\gcphi^{2}\to\sER_{\infty}$
at infinity. This imposes a necessary condition $\left\langle \sER_{\infty}\right\rangle _{0}^{3}=\left\langle \sER_{\infty}\right\rangle _{1}^{3}=0$
for the $\beta_{(n)}$-parameters; for example, requiring $\sER_{\infty}=1$
implies,
\begin{align}
\betap 0+3\betap 1+3\betap 2+\betap 3 & =0,\quad\betap 1+3\betap 2+3\betap 3+\betap 4=0,\label{eq:betaties}
\end{align}
which fixes $\betap 0$ and $\betap 4$ in terms of the other parameters.
Moreover, in spherical symmetry, the metric components must be even
functions of $r$ \cite{Alcubierre:2010is}. Therefore, we use the
Neumann boundary conditions at $r=0$ requiring $\partial_{r}\gcphi=0$
and $\partial_{r}\fcphi=0$. Note that we may still choose some arbitrary
positive values for $\gcphi$ and $\fcphi$ at $r=0$.

\begin{figure}[H]
\noindent \begin{centering}
\begin{tabular}{rrrr}
\panL{1} & \includegraphics{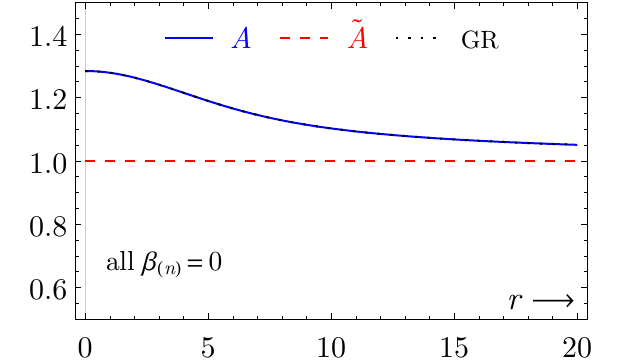} & \panL{2} & \includegraphics{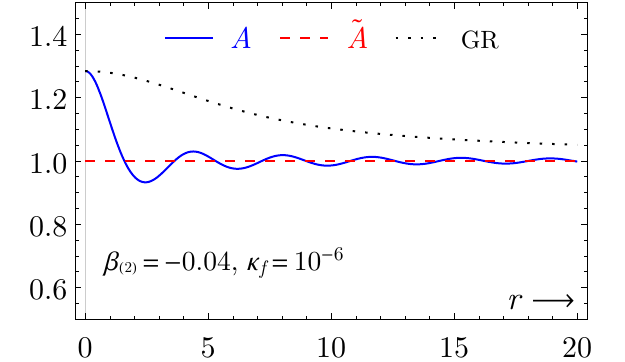}\tabularnewline
\panL{3} & \includegraphics{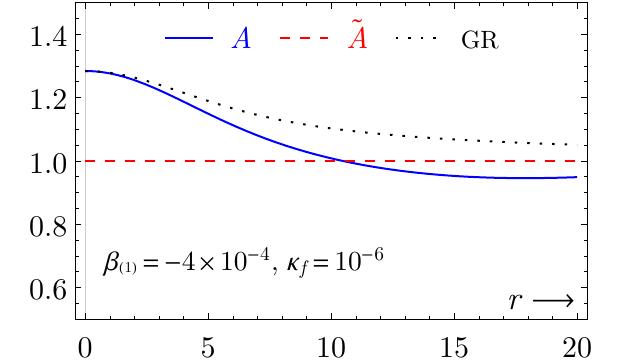} & \panL{4} & \includegraphics{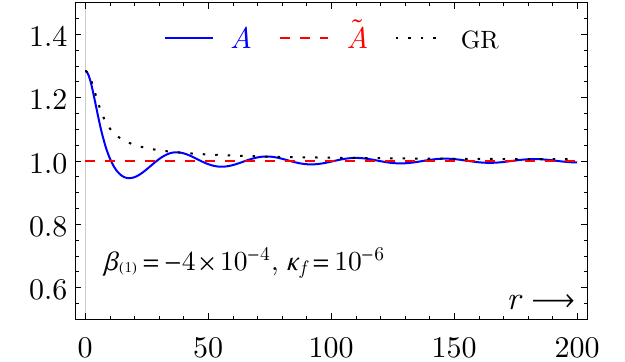}\tabularnewline
\panL{5} & \includegraphics{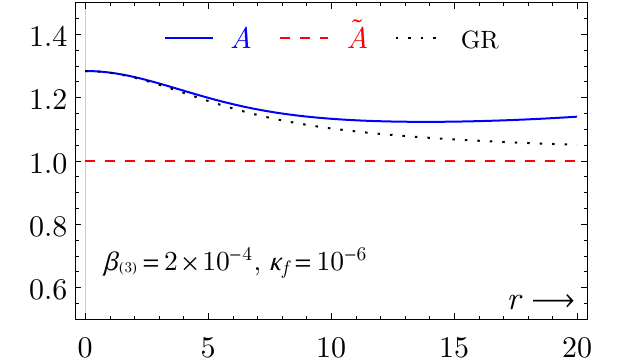} & \panL{6} & \includegraphics{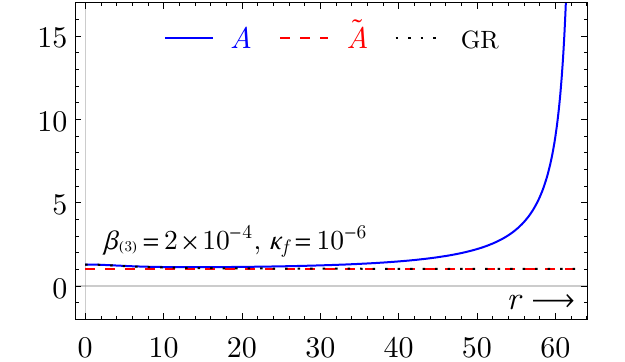}\tabularnewline
\panL{7} & \includegraphics{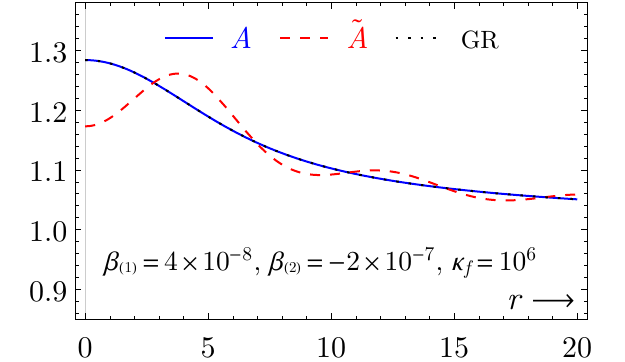} & \panL{8} & \includegraphics{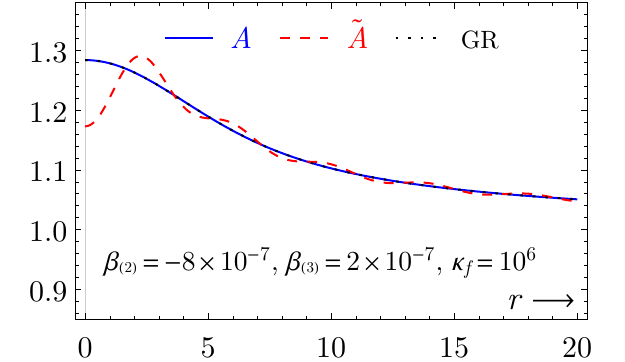}\tabularnewline
~\panL{9} & \includegraphics{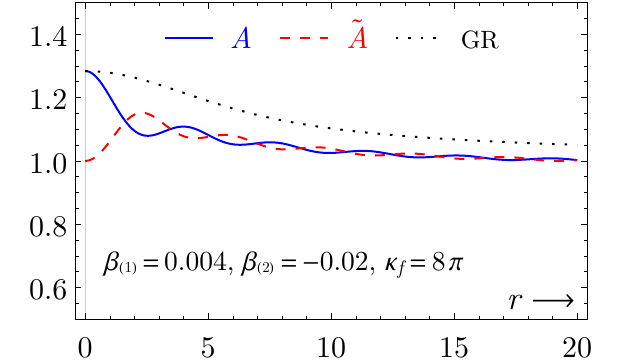} & \panL{10} & \includegraphics{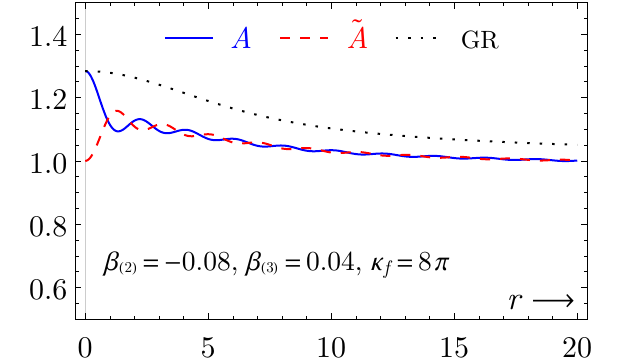}\tabularnewline
\end{tabular}
\par\end{centering}
\caption{\label{fig6:bim-ID}Bimetric polytrope examples: (1) decoupled sectors;
(2) $\protect\fKappa\ll\protect\gKappa$ and $\protect\betap 2=-0.04$;
(3\textendash 4) lower frequency modes for $\protect\betap 2=-4\times10^{-4}$;
(5\textendash 6) initial data diverging at a finite radius for $\protect\betap 2=2\times10^{-4}$;
(7) $\protect\fKappa\gg\protect\gKappa$, $\protect\betap 1=4\times10^{-8}$
and $\protect\betap 2=-2\times10^{-7}$; (8) $\protect\fKappa\gg\protect\gKappa$,
$\protect\betap 2=-8\times10^{-7}$ and $\protect\betap 1=2\times10^{-7}$;
(9) $\protect\betap 1=0.004$ and $\protect\betap 2=-0.02$ for $\protect\fKappa=\protect\gKappa$;
(10) $\protect\betap 2=-0.08$ and $\protect\betap 1=0.04$ for $\protect\fKappa=\protect\gKappa$.
All solutions have $\protect\gKappa=8\pi$. The gravitational constants
are $\protect\fKappa\ll\protect\gKappa$ in (1\textendash 6), $\protect\fKappa\gg\protect\gKappa$
in (7\textendash 8), and $\protect\gKappa=\protect\fKappa$ in (9\textendash 10).
The solved initial data may diverge at finite $r$ for some parameters,
as in (5). Oscillation frequencies are higher for large $\protect\betap n$-parameters.
Note that $\protect\fEA\equiv1$ in (1), but $\protect\fEA\protect\ne1$
in (2\textendash 6). The GR limit is (7\textendash 8) where $\protect\gKappa/\protect\fKappa\to0$
and $\protect\gEA\approx A_{\mathrm{GR}}$.}
\end{figure}

To summarize: beside $\gKappa$, $\fKappa$, $\betap 1$, $\betap 2$,
$\betap 3$, $\betaScale$, and the density $\pfD$, we are free to
choose $\gcphi$ and $\fcphi$ at $r=0$, and their ratio at $r\to\infty$.
Note that the departure from the reference GR solution is a smooth
function of the free parameters (as found by inspection). The GR limit
of the HR theory is given by $\gKappa/\fKappa\to0$ \cite{Schmidt-May:2015vnx,Akrami:2015qga,Berezhiani:2007zf,Martin-Moruno:2013wea}.
Examples of initial data for different parameter values are shown
in figure \ref{fig6:bim-ID}.

\begin{figure}[H]
\noindent \begin{centering}
\hspace{-1em}\includegraphics{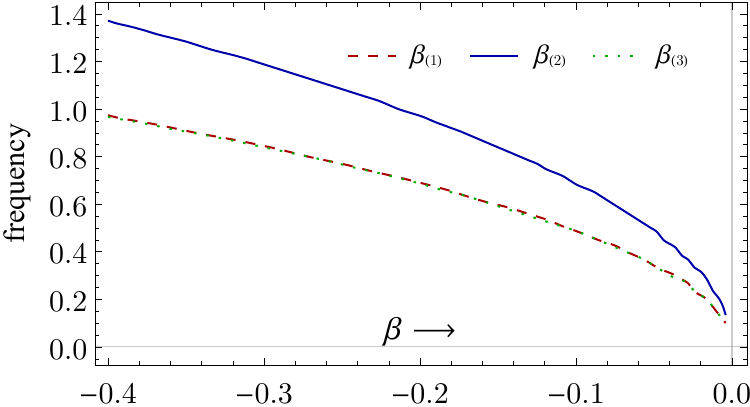}\vspace{-1ex}
\par\end{centering}
\caption{\label{fig7:freqs}Leading frequencies in the oscillations of $\protect\gEA$
at large radii. The values are extracted from the power spectrum of
$\protect\gEA$, each obtained by solving the initial data for a different
$\beta_{(n)}$-model.}
\end{figure}

\noindent From the plots we observe higher oscillation frequencies
for large $\betap n$-parameters, which is compliant with the properties
of the Lane\textendash Emden equation \cite{Chandra:1957intro,Christensen:1994ic}.
The leading frequencies for the pure bimetric polytropes are shown
in figure \ref{fig7:freqs}.

\begin{figure}[H]
\noindent \begin{centering}
\includegraphics{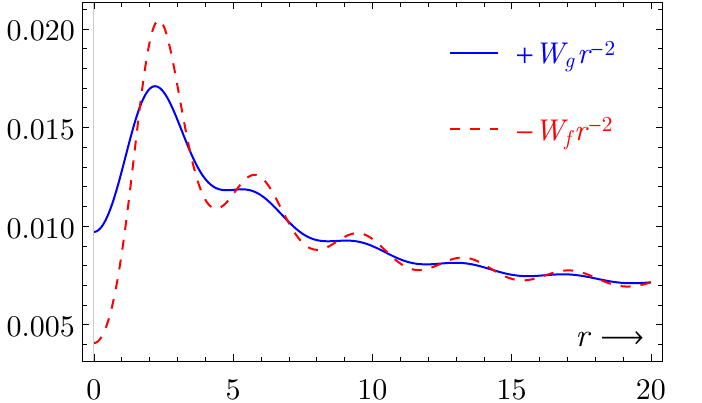}\vspace{-1ex}
\par\end{centering}
\caption{\label{fig8:Ws}Functions $\protect\gW$ and $\protect\fW$ for the
initial data in figure \ref{fig6:bim-ID} panel 9.}
\vspace{-2ex}
\end{figure}

\begin{figure}[H]
\noindent \begin{centering}
~~\includegraphics{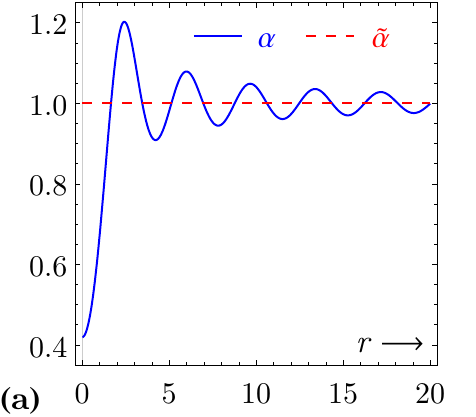}~\includegraphics{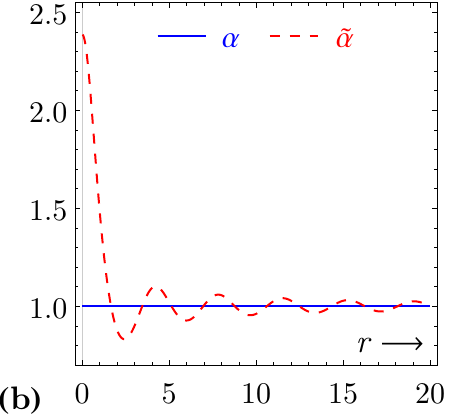}~\includegraphics{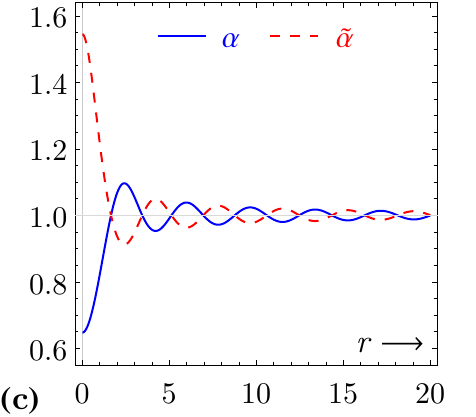}\vspace{-2ex}
\par\end{centering}
\caption{\label{fig9:lapses}The lapses at $t=0$ depending on the gauge (a)
$\protect\fLapse=1$, (b) $\protect\gLapse=1$, and (c) $\protect\gAlpha\protect\fAlpha=\protect\sLt$.}
\end{figure}

The preservation of the additional constraint (\ref{eq:ssym-cc2})
relates the two lapses through $\gAlpha\gW+\fAlpha\fW=0$ where $\gW$
and $\fW$ are functions of dynamical variables \cite{Alexandrov:2012yv,Hassan:2018mbl}.
For the case of spherical symmetry, these functions are established
in \cite{Kocic:2019zdy}. The relation holds at all times, including
$t=0$ where it depends on the initial data as shown in figure \ref{fig8:Ws}
and \ref{fig9:lapses}. 

\noindent Symmetries in the shapes of $\gW$ and $\fW$ at $t\!=\!0$
may be used as a guideline to what kind of gauge conditions are preferable
for the time slicing, e.g., singling out (c) from figure \ref{fig9:lapses}.

%% file: sec-30.tex
\section{Development of the initial data}

\label{sec:evol}

Here we present the solutions obtained from the bimetric initial data.
We use a free evolution scheme where the constraint equations are
solved at the initial surface and the initial data developed by the
bimetric evolution equations. The constraint equations are used as
error estimators during the evolution. To evolve the equations, we
have written a numerical relativity code package for the bimetric
theory, \texttt{bim-solver}. The package is written in C++ and provides
a framework that combines the evolution equations generated in Mathematica
into a functioning bimetric numerical relativity code. The code is
OpenMP parallelized \cite{OpenMP} and supports MPI \cite{MPISpec}.
The implementation details for \texttt{bim-solver} will be given
elsewhere.

\subsection{Formalism and numerical methods}

The evolution equations are of the standard 3+1 form, given in appendix
\ref{app:ev-eqs}. Because of the coordinate singularity at $r=0$,
the equations of motion have to be regularized; for this, we use the
regularization procedure described in \cite{Alcubierre:2004gn,Ruiz:2007rs,Alcubierre:2010is,Alcubierre:2012intro}.
The regularized equations are relegated to an ancillary file (which
can be found on arXiv).

The equations are solved using the finite difference approximation
in a uniform grid. The implemented spatial difference scheme is centered
in space and with arbitrary finite difference order (we have used
up to the sixth order). For the time evolution, we employ the method
of lines (MoL) using Runge\textendash Kutta (RK) and iterated Crank-Nicholson
(ICN) integration. Moreover, Kreiss-Oliger (KO) dissipation has been
added for stability \cite{Kreiss:1973aa,Gustafsson:2013KO}. As diagnostics,
the $l^{2}$-norm of the constraint equations is monitored. 

Boundary conditions should be handled appropriately to avoid spurious
back reflections of waves causing numerical instabilities. As mentioned,
the inner boundaries are  fixed by mirroring the grid cells around
$r=0$. The outer boundaries are handled by imposing open boundary
conditions. We assume that no waves are traveling in from outside
of the domain of influence. Hence, the ghost cells at $r>r_{\mathrm{max}}$
are updated by extrapolating the interior solution, assuming continuity
at the boundary. The extrapolation of the fields is to first order;
hence, small unphysical errors are generated in numerical simulations.
To delay their propagation, we set the right grid boundary at large
radii.

The gauge variables are $\hShift$, $\gAlpha$, and $\fAlpha$ where
the lapses related through $\gAlpha\gW+\fAlpha\fW=0$. The simplest
gauge choices are the algebraic: $\gAlpha=1$ , $\fAlpha=1$, or $\gAlpha\fAlpha=\sLt$.
The last choice fixes to one the lapse function of the geometric mean
metric \cite{Kocic:2018ddp} (other gauges specific to the HR theory
are discussed in \cite{Torsello:meang}). Besides using the algebraic
gauge conditions, the numerical code also implements the maximal slicing,
in this paper used with respect to $\gMet$.

To detect black hole formation, the code implements an apparent horizon
finder in both sectors. In spherical symmetry, a marginally outer
trapped surface (MOTS) is given by \cite{Baumgarte:1996aa,Thornburg:2006zb,Shibata:2015nr},
\begin{equation}
\zeta(r)\coloneqq\partial_{r}(\log\gEB)-\gEA\gK_{2}=0.\label{eq:MOTS_spherical}
\end{equation}
The apparent horizon is the outermost MOTS at which (\ref{eq:MOTS_spherical})
is satisfied. The solution is obtained using a root-finding algorithm.
Alternatively, the function $\zeta(r)$ can be plotted, and the apparent
horizons determined graphically.

%% file: sec-33.tex
\subsection{Simulations}

Here we give the numerical details for the simulations and show representative
results. We have performed two kinds of simulations:
\begin{enumerate}[itemsep=0.5ex,parsep=0pt,label=(\roman*)]
\item benchmark (diagnostics) simulation for the reference GR initial data,
\item simulations with the engaged bimetric interactions.
\end{enumerate}
For the GR simulation, the grid spacing is $\Delta r=0.01$ with the
outer boundary at $r=80$. The spatial difference scheme is of the
fourth order. The integration employs the method of lines with the
third order Runge\textendash Kutta and the Courant\textendash Lax
factor (CFL) of $\Delta t/\Delta r=0.5$. The added Kreiss\textendash Oliger
dissipation is of the fourth order with the coefficient $0.03$. The
lapse is evolved using maximal slicing. The evolution is stable and
goes beyond $t=60$. The results of the GR simulation are shown in
figure \ref{fig5:ref-gr}, which are compatible with \cite{Nakamura:1980}.

\begin{figure}[H]
\noindent \begin{centering}
\vphantom{x}
\par\end{centering}
\noindent \begin{centering}
\begin{tabular}{rrrr}
\panL{a}~ & \includegraphics{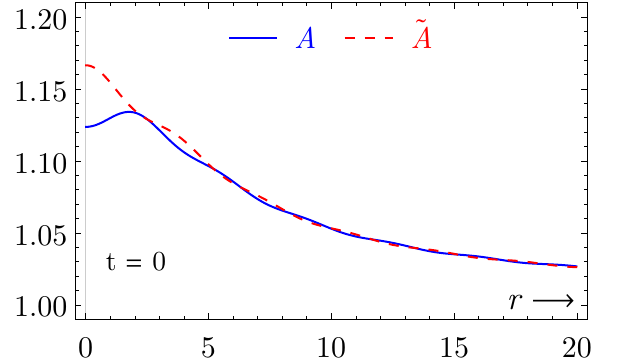} & \panL{b}~ & \includegraphics{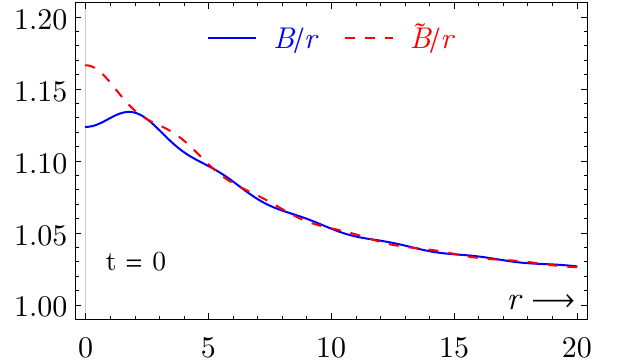}\tabularnewline
\panL{c}~ & \includegraphics{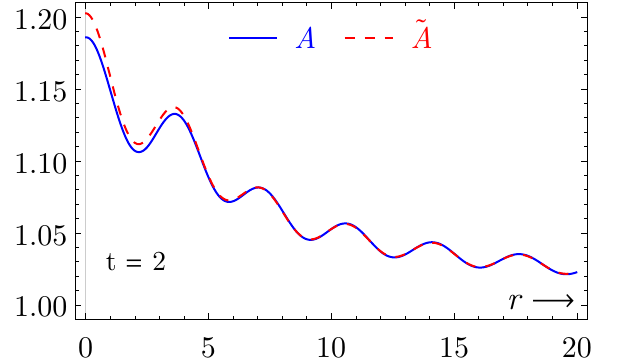} & \panL{d}~ & \includegraphics{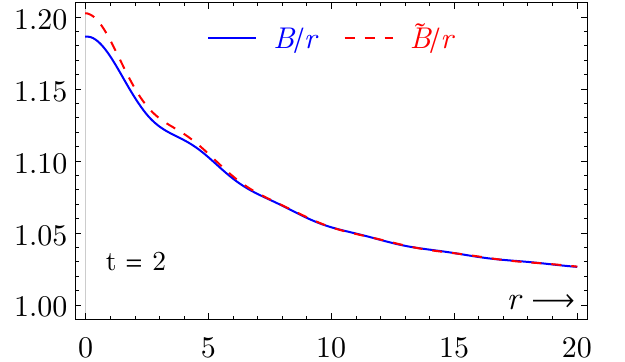}\tabularnewline
\panL{e}~ & \includegraphics{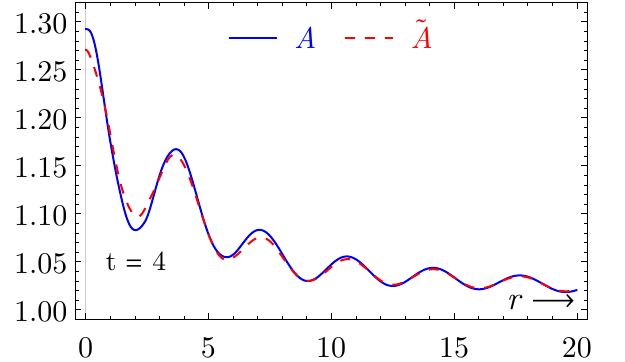} & \panL{f}~ & \includegraphics{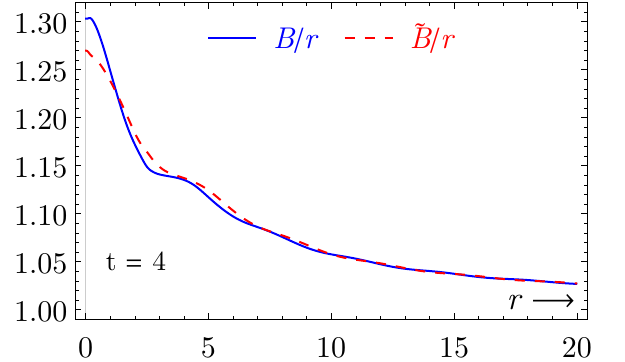}\tabularnewline
\end{tabular}\vspace{-1ex}
\par\end{centering}
\caption{\label{fig10:sim-id}The metric components at $t=0$, $t=2$, and
$t=4$. The initial values are for $\protect\gKappa=\protect\fKappa=8\pi$,
$\protect\betap 1=-1$, $\protect\gcphi\vert_{r=0}=1.06$, $\protect\fcphi\vert_{r=0}=1.08$,
and $\protect\sER_{\infty}=1$. Note that $\protect\gEA=\protect\gEB/r$
and $\protect\fEA=\protect\fEB/r$ at $t=0$ with the tendency $\protect\gEA\approx\protect\fEA$
and $\protect\gEB\approx\protect\fEB$ at later times.}
\end{figure}

After the control benchmark, we consider the development of the bimetric
initial data constructed in subsection \ref{ssec:bimpolytrope}. Here
we use the grid spacing $\Delta r=0.04$ with the outer boundary pushed
to $r=300$ and the CFL decreased to $0.25$. The initial conditions
with the evolution of the metric components are shown in figure \ref{fig10:sim-id}(a).

\begin{figure}[H]
\noindent \begin{centering}
\vphantom{X}
\par\end{centering}
\noindent \begin{centering}
\includegraphics[width=6cm]{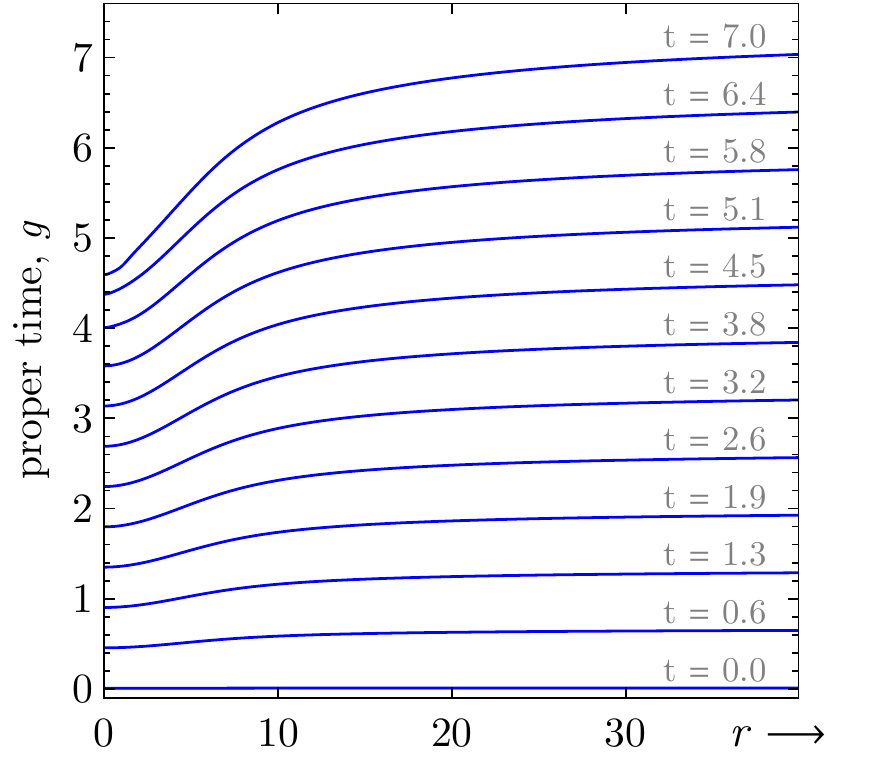}~\includegraphics[width=6cm]{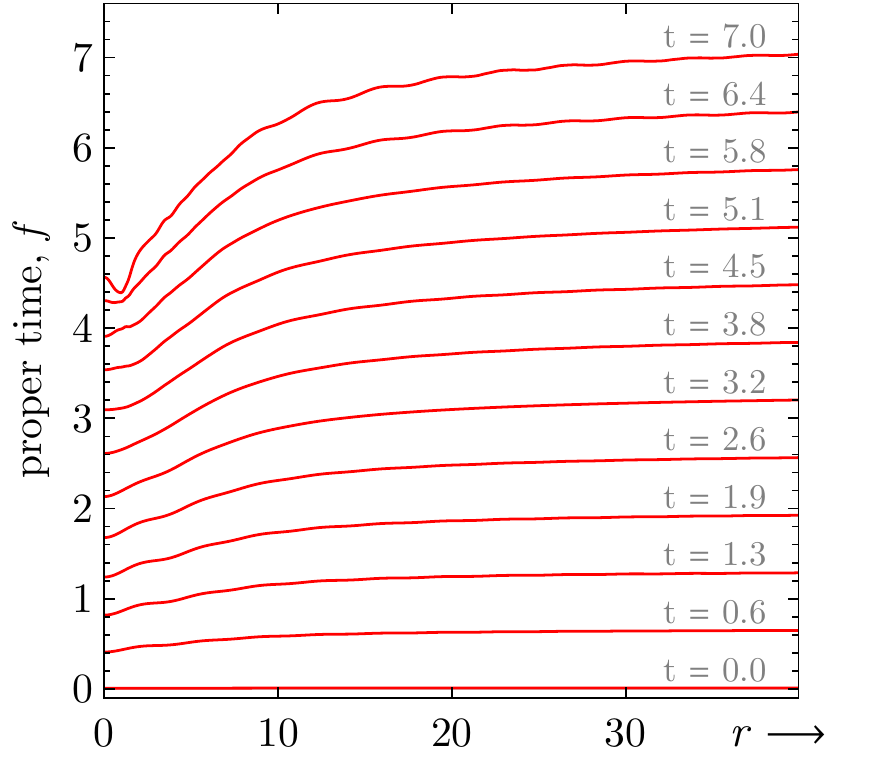}\vspace{-1ex}
\par\end{centering}
\caption{\label{fig11:slicing}Time slicing represented by the proper time
of the Eulerian observers for $\protect\gMet$ and $\protect\fMet$.}
\end{figure}

\begin{figure}[H]
\noindent \begin{centering}
\includegraphics{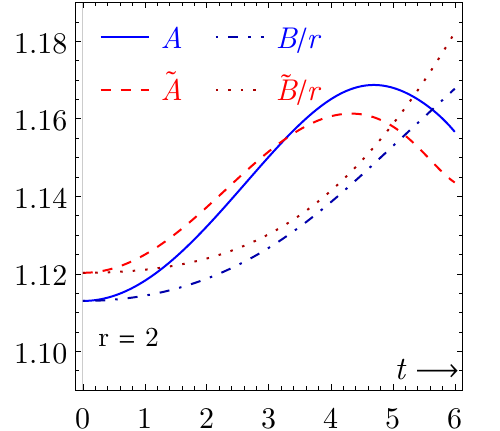}~\includegraphics{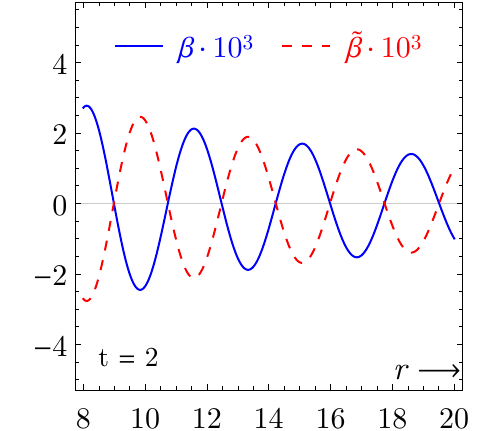}~\includegraphics{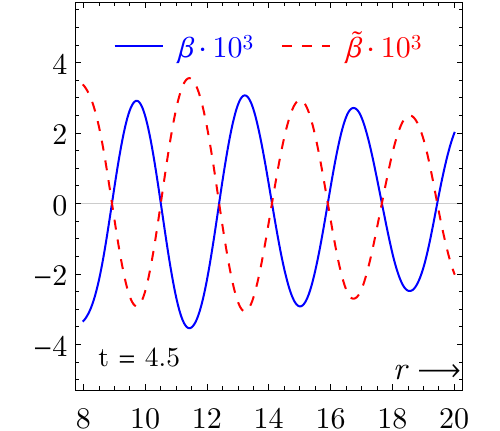}\vspace{-1ex}
\par\end{centering}
\caption{\label{fig12:sim-osc}The time variation of $\protect\gEA$ and $\protect\fEA$
at $r=2$, and the shift vectors at $t=2$ and $t=4.5$.}
\end{figure}

\noindent The metric components have a tendency to become bi-proportional,
$\gEA\approx\fEA$ and $\gEB\approx\fEB$, at least over a short time
period as in figure \ref{fig12:sim-osc}. Note that the null cones
oscillate over the time; the same figure shows the radial components
of the shift vectors at $t=2$ and $t=4.5$.

\begin{figure}[t]
\noindent \begin{centering}
\vphantom{X}
\par\end{centering}
\noindent \begin{centering}
\panL{a}~~\includegraphics[width=12cm]{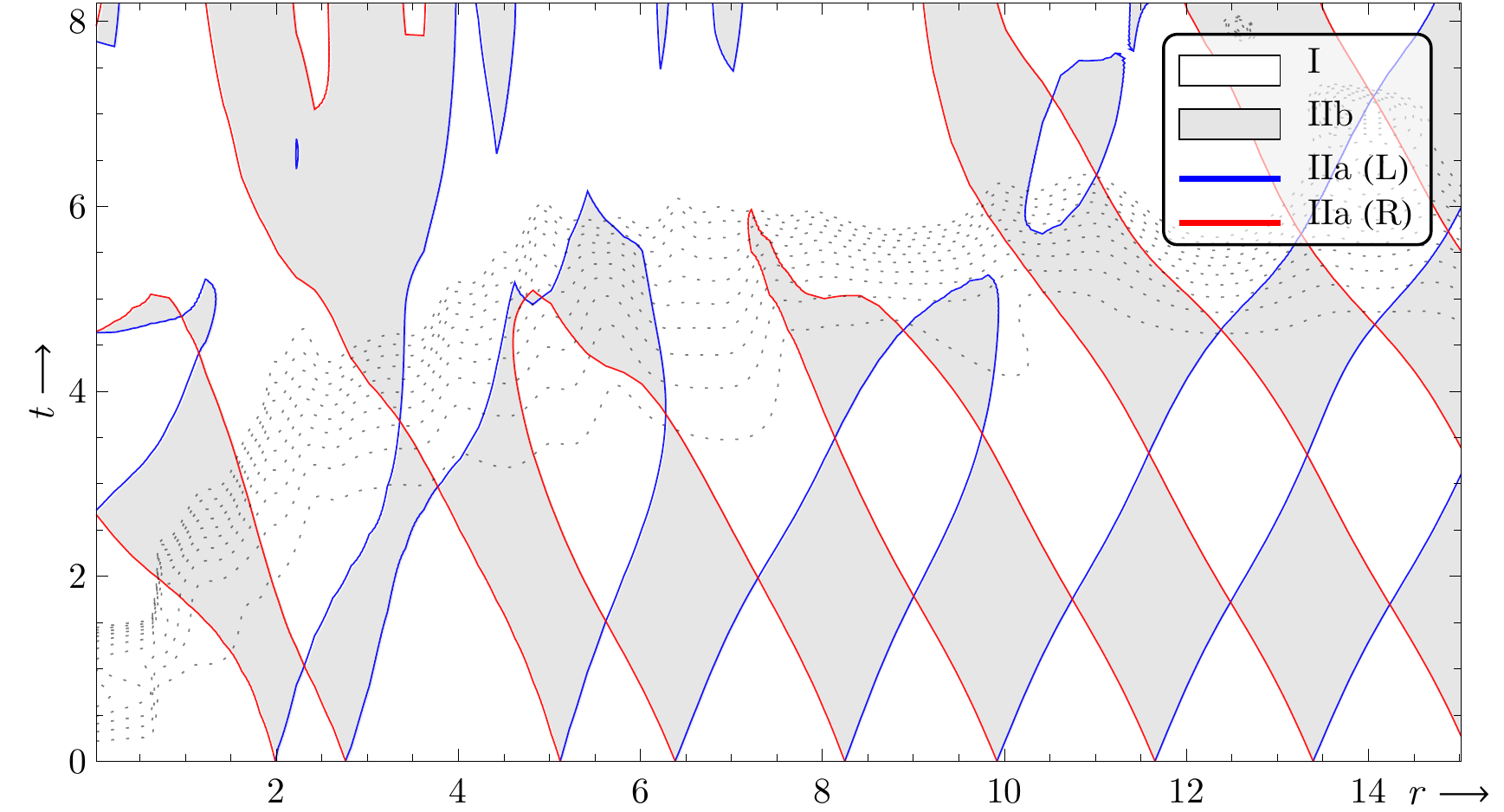}
\par\end{centering}
\noindent \begin{centering}
\panL{b}~~\includegraphics[width=12cm]{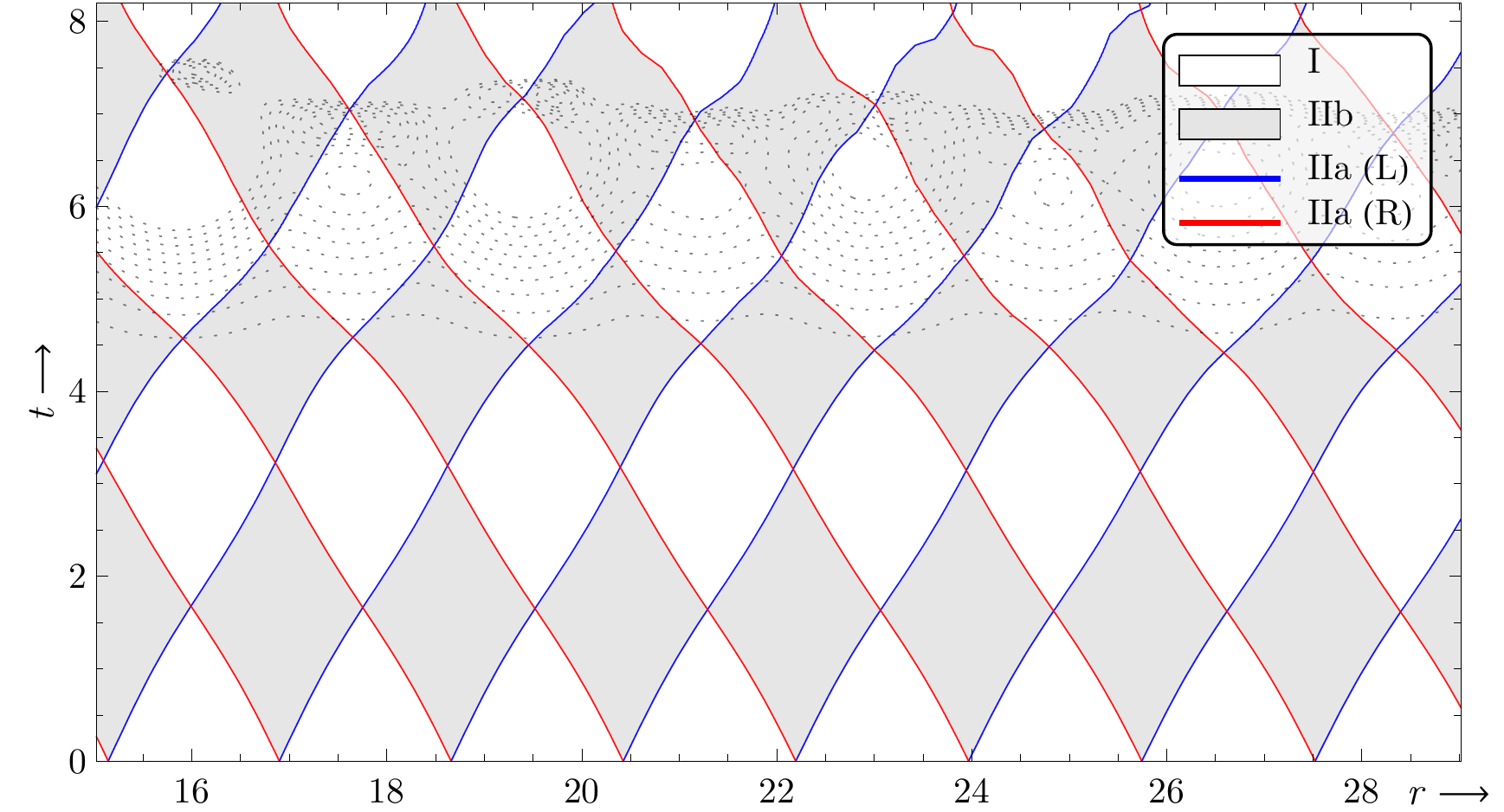}\vspace{-1ex}
\par\end{centering}
\caption{\label{fig13:sim-cd}Causal diagram for the bimetric polytrope: (a)
near $r=0$, and (b) at large radii.}
\end{figure}

\begin{figure}[t]
\noindent \begin{centering}
\includegraphics{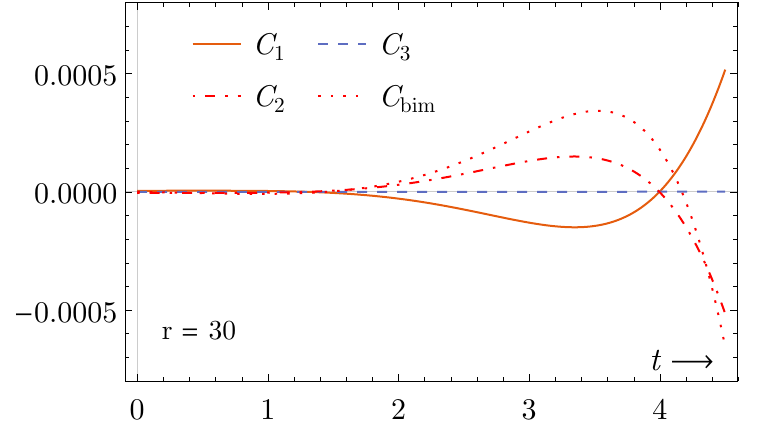}\vspace{-1ex}
\par\end{centering}
\caption{\label{fig14:sim-cerr}The constraint violations at large radii. The
solution is trustworthy for $t<5$.}
\end{figure}

\noindent The lapses are evolved using the maximal slicing with respect
to $\gMet$; the proper times of the Eulerian observers for $\gMet$
and $\fMet$ are shown in figure~\ref{fig11:slicing}. Time is slowed
down near the coordinate origin because of the failed regularization.
In contrast to the GR simulations that run beyond $t>60$, the bimetric
evolutions fail within $t<15$, mostly because of the violated boundary
conditions. The main cause of the instability are irregularities coming
from round-off errors when calculating the lapse ratio and the corresponding
spatial derivatives near $r=0$. The causal diagram for the solution
near the coordinate origin is shown in figure~\ref{fig13:sim-cd}(a).
The dotted lines indicate surface levels of the $l^{2}$-norm of the
scalar constraint violations with separation $10^{-3}$. The irregularities
are visible close to the origin. As noted before, the white regions
are Type I (bidiagonal) and the shaded Type IIb. The edges are Type
IIa, either with the common left- or right-null directions. The intersections
of two left-right Type IIa edges are of Type I. The causal diagram
at larger radii is given in figure~\ref{fig13:sim-cd}(b), with the
constraint violations in figure \ref{fig14:sim-cerr}. The solution
is trustworthy for $t<5$, which is enough to conclude  the time-dependent
nonbidiagonality in the dynamics of the two metrics represented by
the interwoven and oscillating null cones.

\begin{figure}[H]
\noindent \begin{centering}
\hspace{-2mm}\begin{tikzpicture}[x=10mm,y=10mm,node distance=1mm,yscale=1]
  \node[anchor=south east, inner sep=0] at (0,0) { 
    \includegraphics[width=58mm]{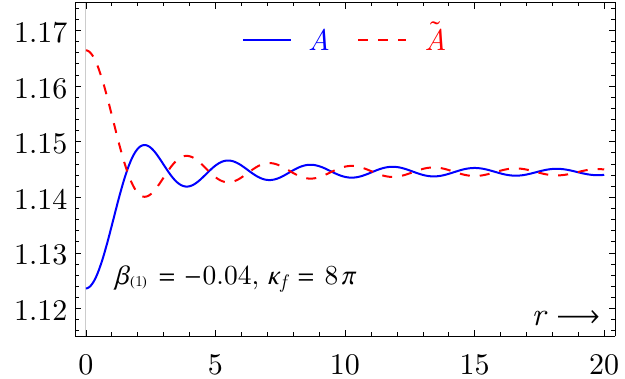}
  };
  \node[anchor=south west, inner sep=0] at (0,0) { 
    \includegraphics[width=96mm]{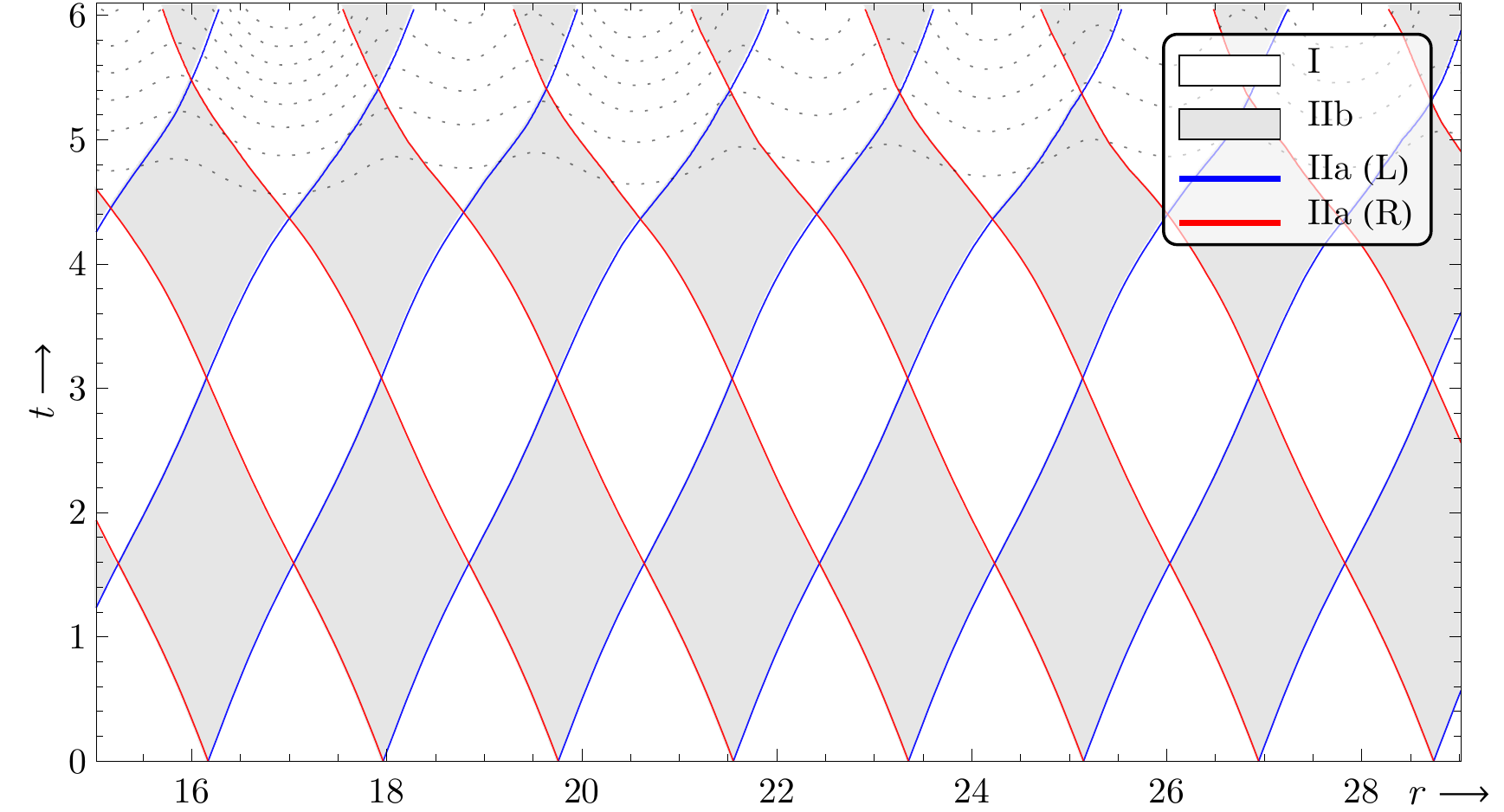}
  };
 \node[] (a) at (-2.6,4.5) 
   {\small Nonstationary vacuum solution:};
 \node[] (a) at (-2.5,2.7) 
   {\footnotesize $(t=0)$};
\end{tikzpicture}\vspace{-1ex}
\par\end{centering}
\caption{\label{fig15:sim-pure}A bimetric polytrope for the same parameters
as earlier but \emph{without matter}, $\protect\pfD=0$. This is a
bimetric nonstationary spherically symmetric solution in vacuum.}
\end{figure}

Finally, we have performed several thousands of simulations for different
parameter values, and the earlier shown examples are typical. An important
observation is that the \emph{pure} bimetric polytropes (vacuum solutions)
share the same property: they are generically nonbidiagonal and nonstationary;
an example is shown in figure \ref{fig15:sim-pure}, where the causal
diagram is trustworthy for $t<5$. The pure bimetric polytropes contradict
the bimetric analogue of Birkhoff's theorem \cite{Birkhoff:1923,Jebsen:1921,Jebsen:2005},
which is compatible with the findings in \cite{Kocic:2017hve}.

\subsection{Gravitational collapse}

\label{3.4}

The simulations are not long enough to shed light on the end point
of gravitational collapse. Nevertheless, the present numerical experiments
show that the collapse of the dust cloud follows the pattern from
the reference GR solution, showing no instabilities. A physically
realistic solution is illustrated in figure \ref{fig17:sim-collapse},
where the initial data are close to GR and $\fKappa\gg\gKappa=8\pi$.
The time variation of the Lagrange shells is shown in panel (b); the
plot is almost the same as in figure \ref{fig5:ref-gr} (the plots
can be compared since the evolution of the time gauge are similar
for the two solutions for $t<7$). 

\begin{figure}[H]
\noindent \begin{centering}
~~\includegraphics{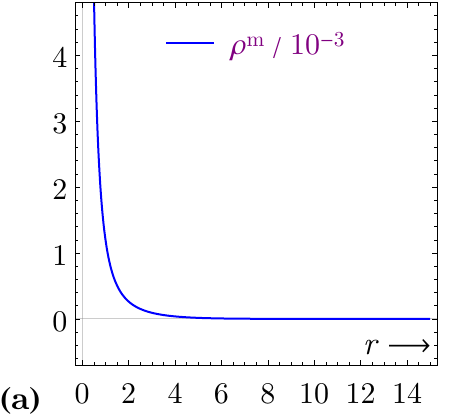}~\includegraphics{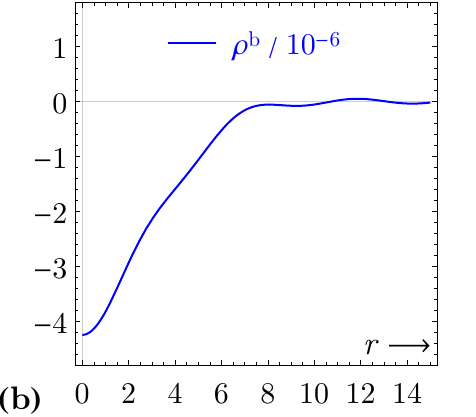}~\includegraphics{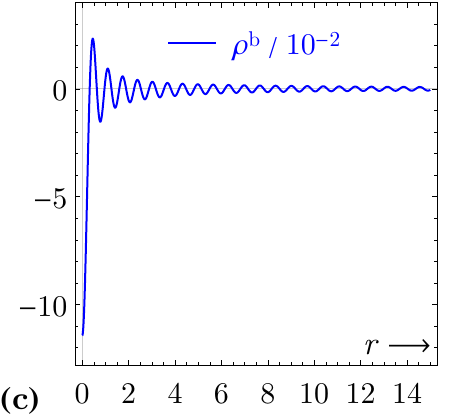}\vspace{-2ex}
\par\end{centering}
\caption{\label{fig16:cmp}The initial stress\textendash energy contributions
coming from: (a) the matter $\protect\grho^{\protect\gmlab}$, (b)
the bimetric potential $\protect\grho^{\protect\gblab}$ in the GR
limit for $\protect\betap 1=-10^{-4}$, $\protect\fKappa=10^{4}$,
and (c) the bimetric potential $\protect\grho^{\protect\gblab}$ when
the initial data are far from GR, having $\protect\betap 1=-1$ and
$\protect\fKappa=8\pi$.}
\end{figure}

\begin{figure}[H]
\noindent \begin{centering}
\hspace{-2mm}\begin{tikzpicture}[x=1mm,y=1mm,node distance=1mm,yscale=1]
  \node[anchor=south east, inner sep=0] at (-10,5) { 
    \includegraphics[width=65mm]{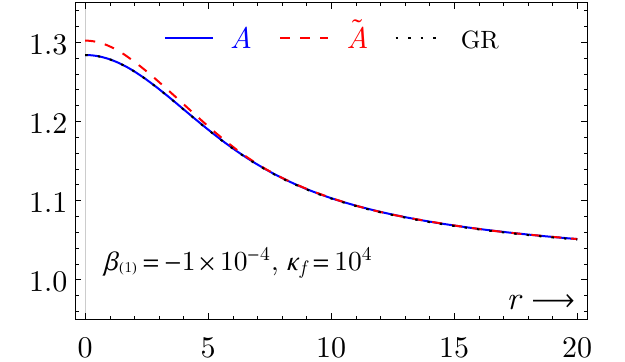}
  };
  \node[anchor=south west, inner sep=0] at (-6,0) { 
    \includegraphics[width=80mm]{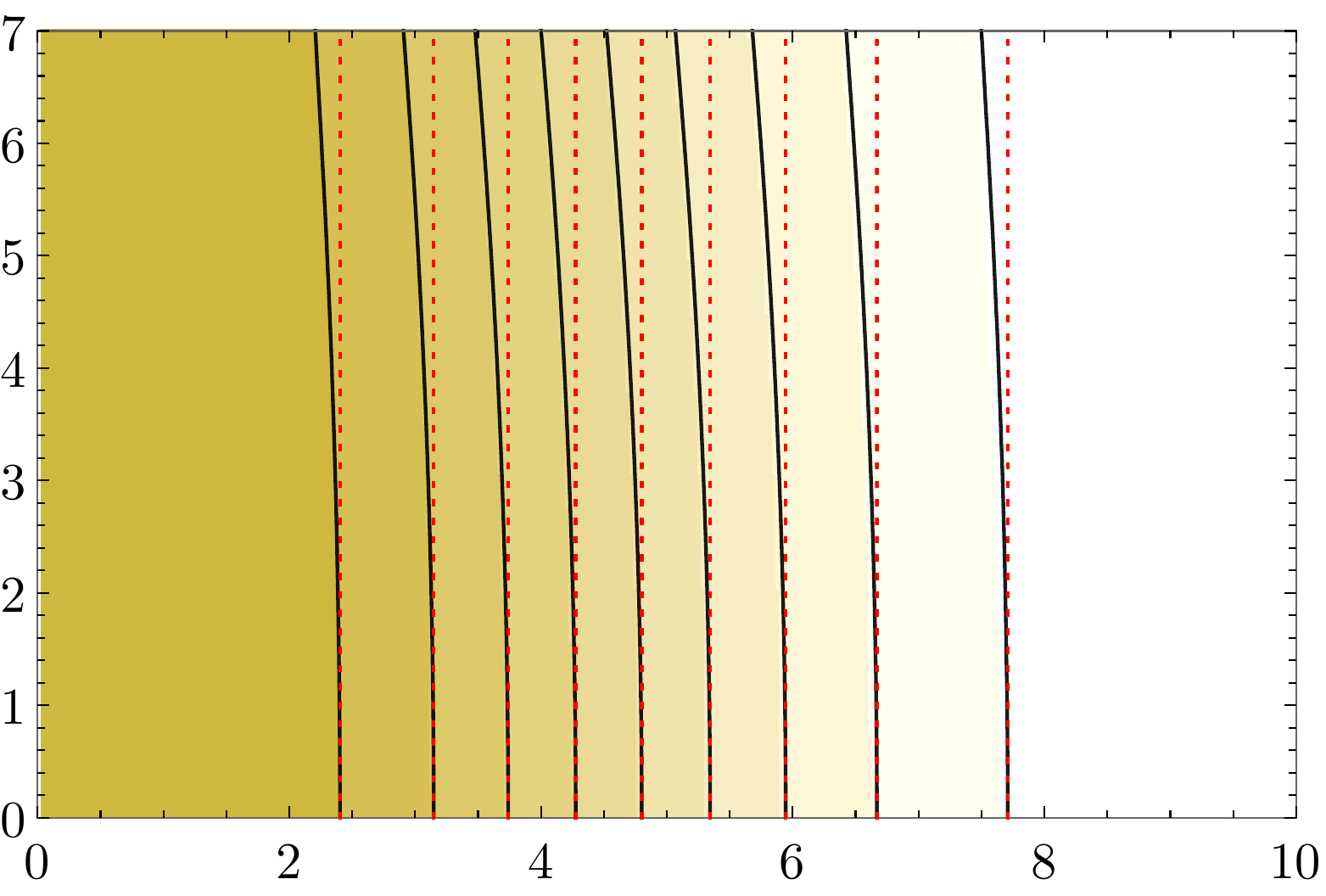}
  };
  \node[anchor=north, inner sep=0] at (3,0) { 
    \includegraphics[width=120mm]{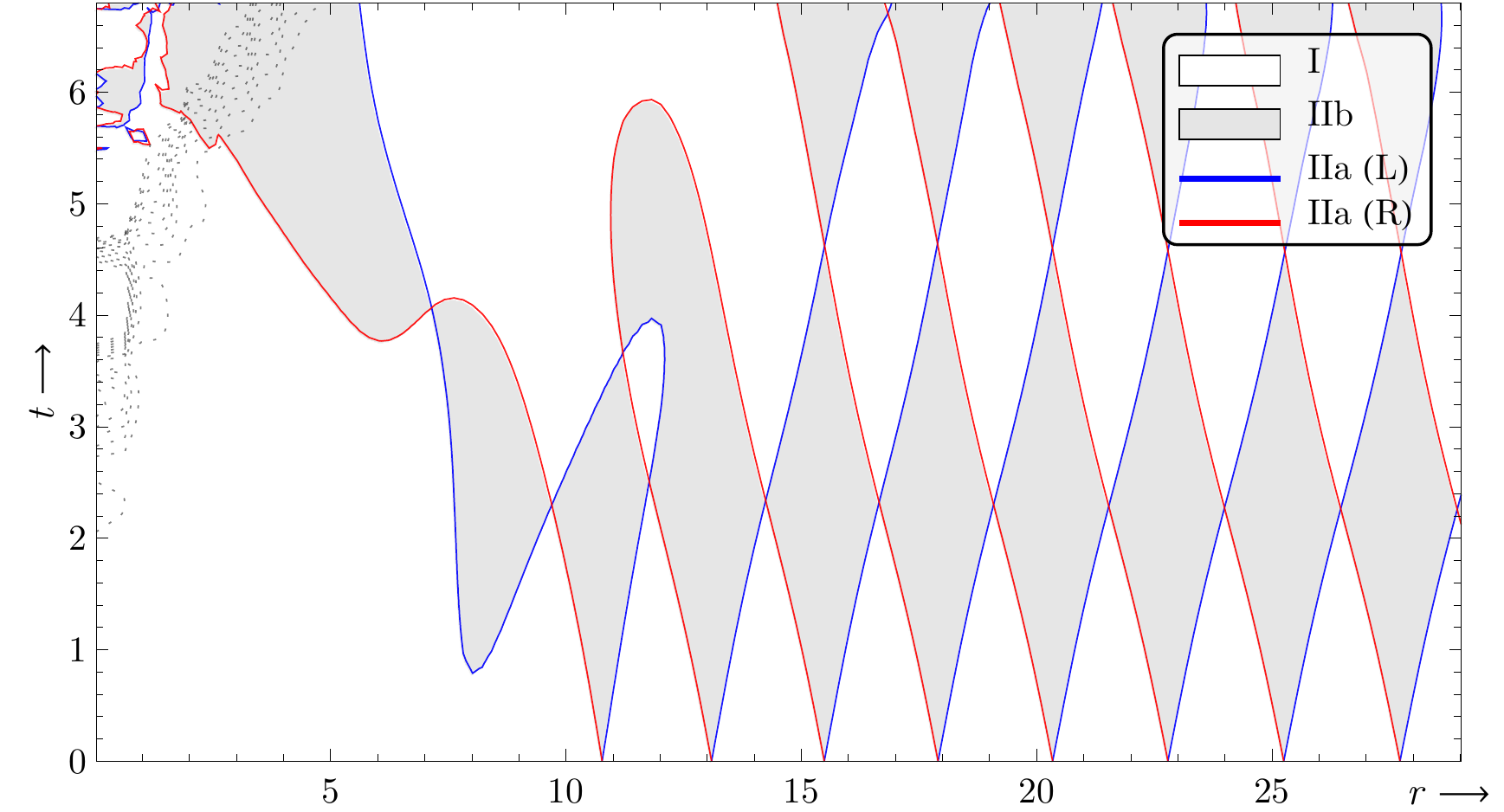}
  };

  \draw [-latex] (-1,8) -- (-1,18); 
  \node [anchor=south west] at (-1.5,16.5) {\small $t$};
  \draw [-latex] (-1,8) -- (9,8);   
  \node [anchor=south west] at (7.5,7.5) {\small $r$};   

  \node [text width=20mm] at (67,40) {\footnotesize Lagrange\\shells\\~};

  \node at (-76,34) {\small\bf (a)};
  \node at (-10,49) {\small\bf (b)};
  \node at (-57,-14) {\small\bf (c)};
\end{tikzpicture}\vspace{-2ex}
\par\end{centering}
\caption{\label{fig17:sim-collapse}(a) A bimetric polytrope where the initial
data are close to the reference GR solution. (b) The gravitational
collapse follows the same pattern as in figure \ref{fig5:ref-gr}.
The vertical dotted lines indicate the initial positions of the Lagrange
shells. (c) Although $\protect\gEA\approx A_{\mathrm{GR}}$, the causal
diagram shows the oscillations in the metric fields. The solution
is trustworthy for $t<7$.}
\end{figure}
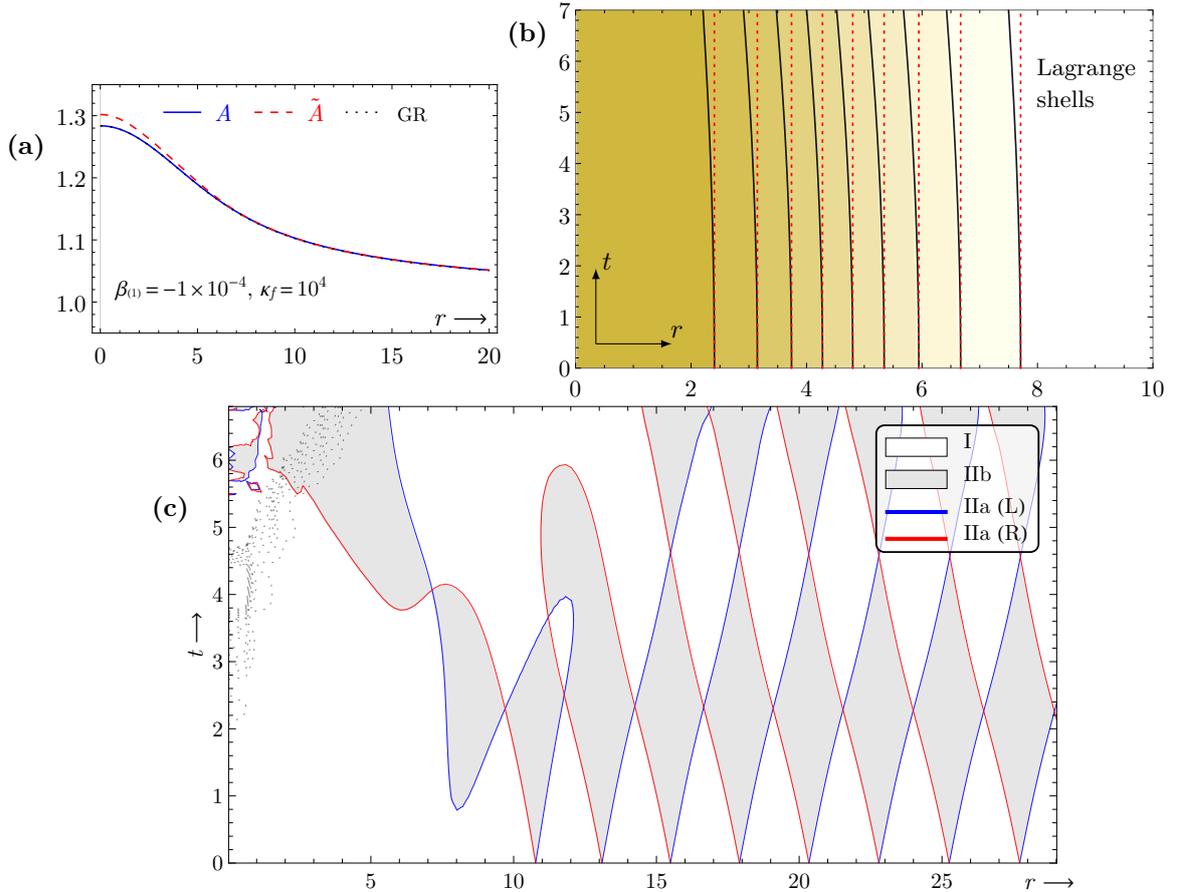

The stress\textendash energy contributions coming from the matter,
$\grho^{\gmlab}$, and the bimetric potential, $\grho^{\gblab}$,
are shown in figures \ref{fig16:cmp}(a) and (b). These contributions
enter the constraint equations (\ref{eq:scCg}) and (\ref{eq:ideqbim1}).
In the GR limit, $\grho^{\gblab}$ is more than $10^{3}$ times smaller
than $\grho^{\gmlab}$, due to small $\beta_{(n)}$-parameters. For
comparison, figure \ref{fig16:cmp}(c) shows a large contribution
from the bimetric potential when the initial data are far from GR
(as in figure \ref{fig2:bim-ex}).

%% file: sec-60.tex
\section{Discussion and outlook}

\label{sec:discussion}

In this work, we have:
\begin{enumerate}[itemsep=0.7ex,parsep=0pt,label=(\roman*)]
\item presented a method for solving the constraint equations in the Hassan\textendash Rosen
theory to determine bimetric initial data by deforming the existing
GR initial data,
\item obtained the generalized Lane\textendash Emden equations for the bimetric
initial data specifically assuming the conformally flat spatial metrics
at the moment of time symmetry,
\item solved for and analyzed the obtained initial data,
\item evolved the initial data using a new numerical code for bimetric relativity.
\end{enumerate}
The results once again point out the importance of nonbidiagonality
in the dynamics of the two metrics. The authors in \cite{Babichev:2014oua}
showed that to have dynamically stable solutions, nondiagonal metric
elements are needed. Moreover, \cite{Torsello:2017cmz} concludes
that considering static black hole solutions in the HR theory is insufficient
and that a full dynamical treatment needs to be performed in order
to investigate the end point of gravitational collapse. Finally, even
in the absence of external matter sources, the bimetric solutions
are essentially nonbidiagonal and nonstationary; hence, our results
disprove the analogue version of Birkhoff's theorem for bimetric theory,
which is compatible with the findings in \cite{Kocic:2017hve}.

The initial data for bimetric polytropes requires negative $\beta_{(n)}$-parameters
because of the boundary conditions that are imposed on the constraint
equations (\ref{eq:ideqbim}) (see figure~\ref{fig7:freqs}). This
would correspond to an imaginary Fierz\textendash Pauli mass obtained
from the spectrum of linearized mass eigenstates on the proportional
backgrounds \cite{Hassan:2012wr,Schmidt-May:2015vnx,Luben:2018ekw}.
However, the bimetric polytropes are nonproportional solutions in
a strong-field regime with the initial conditions that assume conformal
flatness. In addition, we are more restrictive on the choice of $\betap 0$
and $\betap 4$ in (\ref{eq:betaties}) than needed to define the
Fierz\textendash Pauli mass, which could affect the conclusions on
its sign. This suggests that the properties of the bimetric parameter
space should be further investigated.

A future goal is to obtain a long term development of the initial
data. For this purpose, the covariant BSSN formalism is established
in \cite{Torsello:2019tgc}, and various gauge conditions specific
to bimetric relativity are investigated in \cite{Torsello:meang}.
Work in progress deals with the implementation of BSSN in \texttt{bim-solver}
to obtain stable numerical simulations in spherical symmetry. 

\subsubsection*{Acknowledgments}

We would like to thank Giovanni Camelio for a careful reading of the
manuscript. \medskip

%% file: sec-90.tex
\section{Evolution equations}

\label{app:ev-eqs}

The evolution equations for the spatial metrics are,\bSe\label{eq:ssym-evol-1}
\begin{alignat}{2}
\partial_{t}\gEA & =\usignK\gAlpha\gEA\gK_{1}+\partial_{r}(\hShift\gEA+\gAlpha\sLv), & \qquad\partial_{t}\gEB & =\usignK\gAlpha\gEB\gK_{2}+\big(\hShift+\gAlpha\gEA^{-1}\sLv\big)\partial_{r}\gEB,\\
\partial_{t}\fEA & =\usignK\fAlpha\fEA\fK_{1}+\partial_{r}(\hShift\fEA-\fAlpha\sLv), & \partial_{t}\fEB & =\usignK\fAlpha\fEB\fK_{2}+\big(\hShift-\fAlpha\fEA^{-1}\sLv\big)\partial_{r}\fEB.
\end{alignat}
\eSe The evolution equations for the extrinsic curvatures are,\bSe\label{eq:ssym-evol-2}
\begin{align}
\partial_{t}\gK_{1} & =\big(\hShift+\gAlpha\gEA^{-1}\sLv\big)\partial_{r}\gK_{1}\isignK\gAlpha\gK_{1}\big(\gK_{1}+2\gK_{2}\big)\signK\gAlpha\gKappa\Big\{\,\gJota_{1}-\frac{1}{2}(\gJota-\grho)\,\Big\}\nonumber \\
 & \qquad\isignK\bigg(\frac{\partial_{r}\gAlpha}{\gEA^{2}}\frac{\partial_{r}\gEA}{\gEA}-\frac{\partial_{r}^{2}\gAlpha}{\gEA^{2}}+2\frac{\gAlpha}{\gEA^{2}}\frac{\partial_{r}\gEA}{\gEA}\frac{\partial_{r}\gEB}{\gEB}-2\frac{\gAlpha}{\gEA^{2}}\frac{\partial_{r}^{2}\gEB}{\gEB}\bigg),\\
\partial_{t}\fK_{1} & =\big(\hShift-\fAlpha\fEA^{-1}\sLv\big)\partial_{r}\fK_{1}\isignK\fAlpha\fK_{1}\big(\fK_{1}+2\fK_{2}\big)\signK\fAlpha\fKappa\Big\{\,\fJota_{1}-\frac{1}{2}(\fJota-\frho)\,\Big\}\nonumber \\
 & \qquad\isignK\bigg(\frac{\partial_{r}\fAlpha}{\fEA^{2}}\frac{\partial_{r}\fEA}{\fEA}-\frac{\partial_{r}^{2}\fAlpha}{\fEA^{2}}+2\frac{\fAlpha}{\fEA^{2}}\frac{\partial_{r}\fEA}{\fEA}\frac{\partial_{r}\fEB}{\fEB}-2\frac{\fAlpha}{\fEA^{2}}\frac{\partial_{r}^{2}\fEB}{\fEB}\bigg),\\
\partial_{t}\gK_{2} & =\big(\hShift+\gAlpha\gEA^{-1}\sLv\big)\partial_{r}\gK_{2}\isignK\gAlpha\gK_{2}\big(\gK_{1}+2\gK_{2}\big)\signK\gAlpha\gKappa\Big\{\,\gJota_{2}-\frac{1}{2}(\gJota-\grho)\,\Big\}\nonumber \\
 & \qquad\isignK\bigg(\frac{\gAlpha}{\gEB^{2}}-\frac{\partial_{r}\gAlpha}{\gEA^{2}}\frac{\partial_{r}\gEB}{\gEB}+\frac{\gAlpha}{\gEA^{2}}\frac{\partial_{r}\gEA}{\gEA}\frac{\partial_{r}\gEB}{\gEB}-\frac{\gAlpha}{\gEA^{2}}\frac{(\partial_{r}\gEB)^{2}}{\gEB^{2}}-\frac{\gAlpha}{\gEA^{2}}\frac{\partial_{r}^{2}\gEB}{\gEB}\bigg),\\
\partial_{t}\fK_{2} & =\big(\hShift-\fAlpha\fEA^{-1}\sLv\big)\partial_{r}\fK_{2}\isignK\fAlpha\fK_{2}\big(\fK_{1}+2\fK_{2}\big)\signK\fAlpha\fKappa\Big\{\,\fJota_{2}-\frac{1}{2}(\fJota-\frho)\,\Big\}\nonumber \\
 & \qquad\isignK\bigg(\frac{\fAlpha}{\fEB^{2}}-\frac{\partial_{r}\fAlpha}{\fEA^{2}}\frac{\partial_{r}\fEB}{\fEB}+\frac{\fAlpha}{\fEA^{2}}\frac{\partial_{r}\fEA}{\fEA}\frac{\partial_{r}\fEB}{\fEB}-\frac{\fAlpha}{\fEA^{2}}\frac{(\partial_{r}\fEB)^{2}}{\fEB^{2}}-\frac{\fAlpha}{\fEA^{2}}\frac{\partial_{r}^{2}\fEB}{\fEB}\bigg).
\end{align}
\eSe The above set of equations is in each sector equivalent to the
GR case \cite{Shibata:2015nr}. 

These equations are subject to the regularization procedure described
in \cite{Alcubierre:2004gn,Ruiz:2007rs,Alcubierre:2010is,Alcubierre:2012intro}.
There are two types of regularity conditions that must be satisfied.
First, the parity which comes from symmetry considerations. Spherical
symmetry implies that a reflection in the radial coordinate leaves
metrics unchanged. After reflecting $r\to-r$, we see that the lapses,
the spatial metric components, and the corresponding components of
the extrinsic curvature must be even functions of $r$, while the
shift vector must be odd. Beside the parity, the extra regularity
conditions require that the manifold must be locally flat at the origin.
The regularization is done in several steps (done in both sectors): 
\begin{enumerate}[itemsep=0.5ex,parsep=0ex,leftmargin=4em,label=(\roman*)]
\item redefine $\gEB\to\gEB r$ (and similarly $\fEB\to\fEB r$),
\item introduce:

\qquad$D_{\gLapse}\coloneqq\partial_{r}\log\gAlpha$, $D_{\gEA}\coloneqq\partial_{r}\log\gEA$,
$D_{\gEB}\coloneqq\partial_{r}\log\gEB$,

\qquad$\sigma=(\gEA-\gEB)/r$, and $\gK_{\Delta}=(\gK_{1}-\gK_{2})/r$,
\item find the evolution equation for these variables,
\item evolve these variables imposing the odd parity conditions.
\end{enumerate}
The regularized equations are too long to write down here; they are
relegated to an ancillary file, which can be found on arXiv.